\documentclass[aoas]{imsart}

\RequirePackage{amsthm,amsmath,amsfonts,amssymb}
\RequirePackage[authoryear]{natbib}
\RequirePackage[colorlinks,citecolor=blue,urlcolor=blue]{hyperref}
\RequirePackage{graphicx}
\usepackage{algorithm}
\usepackage{bm,bbm}

\graphicspath{ {plots/} }

\startlocaldefs
\newcommand{\vect}[1]{\boldsymbol{\mathbf{#1}}}
\newcommand{\indicator}{\mathbbm{1}}
\newcommand{\red}[1]{\textcolor{black}{#1}}

\newcommand{\reals}{\mathbb{R}}
\newcommand{\ybf}{\boldsymbol{y}}
\newcommand{\Ybf}{\boldsymbol{Y}}
\newcommand{\Lbf}{\boldsymbol{L}}
\newcommand{\xbf}{\boldsymbol{x}}
\newcommand{\Xbf}{\boldsymbol{X}}
\newcommand{\Sbf}{\boldsymbol{S}}
\newcommand{\etabf}{\boldsymbol{\eta}}
\newcommand{\Lambdabf}{\boldsymbol{\Lambda}}
\newcommand{\Thetabf}{\boldsymbol{\Theta}}

\newcommand{\Gammabf}{\boldsymbol{\Gamma}}
\newcommand{\gammabf}{\boldsymbol{\gamma}}
\newcommand{\lambdabf}{\boldsymbol{\lambda}}
\newcommand{\epsilonbf}{\boldsymbol{\epsilon}}
\newcommand{\Dbf}{\boldsymbol{D}}
\newcommand{\Tbf}{\boldsymbol{T}}
\newcommand{\eye}{\boldsymbol{I}}
\newcommand{\Sigmabf}{\boldsymbol{\Sigma}}
\newcommand{\kron}{\!\otimes\!}
\newcommand{\one}{\vect{1}}

\endlocaldefs

\begin{document}

\begin{frontmatter}
\title{
Feature aware covariance estimation, with application to mixtures of chemical exposures
}
\runtitle{Feature aware covariance estimation}

\begin{aug}
\author[A]{\fnms{Elizabeth}~\snm{Bersson}\ead[label=e1]{ebersson@mit.edu}},
\author[B]{\fnms{Kate}~\snm{Hoffman}\ead[label=e2]{kate.hoffman@duke.edu}},
\author[B]{\fnms{Heather M.}~\snm{Stapleton}\ead[label=e4]{heather.stapleton@duke.edu}},
\and
\author[C]{\fnms{David B.}~\snm{Dunson}\ead[label=e3]{dunson@duke.edu}}
\address[A]{Laboratory for Information and Decision Systems, Massachusetts Institute of Technology\printead[presep={,\ }]{e1}}

\address[B]{Nicholas School of the Environment, Duke University\printead[presep={,\ }]{e2,e4}}

\address[C]{Department of Statistical Science, Duke University\printead[presep={,\ }]{e3}}
\end{aug}

\begin{abstract}
The motivation of this article is to improve inferences on the covariation in environmental exposures, motivated by data from a study of Toddlers Exposure to SVOCs in Indoor Environments (TESIE). The challenge is that the sample size is limited, so empirical covariance provides a poor estimate. In related applications, Bayesian factor models have been popular; these approaches express the covariance as low rank plus diagonal and can infer the number of factors adaptively. However, they have the disadvantage of shrinking towards a diagonal covariance, often under estimating important covariation patterns in the data. Alternatively, the dimensionality problem is addressed by collapsing the detailed exposure data within chemical classes, potentially obscuring important information. We apply a \red{feature aware covariance} regression extension of Bayesian factor analysis, which improves performance by including information from features summarizing properties of the different exposures. 
This approach enables shrinkage to more flexible covariance structures, reducing the over-shrinkage problem, as we illustrate in the TESIE data using various 
chemical features.
\end{abstract}


\end{frontmatter}


\section{Introduction}


Many environmental health studies are conducted that gather measurements
of chemical exposures from multiple exposure pathways, often within a relatively
small pool of subjects. In general, the chemicals targeted in these studies are suspected to
have an impact on health, and understanding the health risks of these simultaneous exposures, or mixtures,
is a key objective of public health initiatives
\citep[e.g., ][]{NIEHS2012}.
Due to synergistic
and antagonistic interaction effects, the impact of chemical exposures varies depending on the mixture arrangement, and, as such, understanding patterns of variation among exposures
is pertinent. In this work, we aim to improve the accuracy of covariance estimation
for exposures measured in the study of Toddlers Exposure to SVOCs in Indoor Environments
(TESIE)
\citep{Hoffman2018}.

Semi-volatile organic compounds (SVOCs) are widely used in everyday consumer
products such as construction materials and household items, including furniture and cleaning products, and personal care products such as nail polish and shampoo. Although the health effects resulting from mixtures of exposures are not yet fully understood, there exists a
wide array of accessible auxiliary information regarding the individual chemicals. In general, for example, chemical properties such as vapor pressure and aqueous solubility are
readily available publicly from reputable sources, including the US National Institutes of Health
\citep{Kim2023} and the US Environmental Protection Agency \citep{CompTox}. 
For the focus chemicals in the TESIE study, much auxiliary
information is known, including chemical class, common use cases, and chemical molecular
properties. Moreover, each vector of exposure measurements has an implicit
matrix structure consisting of exposure measurement tool by chemical, information that would ideally be incorporated into a statistical analysis. 
In fact, in practice, this auxiliary information is often used by experts to intuit or describe some scientific or practical differences between exposures. In this work, we formalize this practice
by using these auxiliary data, or \red{outcome features}, to improve covariance estimation
accuracy by incorporating them into a modeling framework.
\red{Formally, we define \textit{outcome features} 
to be predictors or covariates that vary with the outcome index instead of the sample index.}

In general, accurate covariance estimation is a challenging objective in high-dimensional settings. For example, a naive
approach estimates a population covariance matrix with the sample covariance matrix.
However, unless the sample size is appreciably larger than the covariance dimension, the sample covariance matrix
will be unstable
\citep{Johnstone2001}.
Whats more, the sample covariance matrix will not be invertible if there are more variables than the number of samples.
This is problematic as the inverse is required for many statistical tasks, including, for example,
classification \citep{Friedman1989} and hypothesis testing \citep[\S 5.3]{Mardia1979}.

For these reasons, among others,
analyzing all available data jointly in an exposure mixture analysis may be infeasible with naive statistical methods.
Consequently, in practice, scientists will implicitly use \red{information from outcome features} to partition the variables in a dataset into subsets 
to be analyzed separately. 
It is common, for example, to analyze exposure mixture data
distinctly 
for each chemical class, as in, for example, \cite{Liu2023,JamesTodd2017}.
With a Gaussian sampling model assumption, this corresponds to imposing an assumption of zero correlation across classes conditional on class-specific modeling assumptions, or, a block diagonal covariance matrix for all variables. More comprehensive approaches have been developed that utilize the outcome feature information that describes distinct categorical groupings of variables
\citep{Bao2024}.

In analyzing exposure mixtures, 
latent factor models are increasingly utilized \citep{Ferrari2021,Roy2021}.
A latent factor model decomposes an unstructured covariance matrix as the sum of a diagonal variance matrix and a possibly low-rank covariance matrix. 
Due to correlation among exposures,
a small number of factors may represent the covariance well, which yields a reduction in the number of unknown parameters to be estimated and, in turn, improved precision.
In addition, this decomposition allows inversion of the covariance matrix in a regime with more variables than samples. 
In a Bayesian framework, state-of-the-art methods have been developed that allow for \red{uncertainty of} the dimension of the low-rank covariance term.
Some examples include \cite{Fruhwirth2025,Legramanti2020,Bhattacharya2011}. 
\red{These approaches assume exchangeability with respect to the order of the feature variables. Moreover, they have a mean of zero and hence} 
often shrink the factor loading matrix  strongly towards a zero-matrix. 
This effectively shrinks the covariance matrix towards a diagonal structure, which may not represent the population well.

Recently, \red{non-exchangeable} approaches have been developed that use outcome feature data to inform the sparsity structure of a covariance matrix in a latent factor model framework \citep{Schiavon2022} and of a precision matrix in a graphical modeling framework \citep{Xi2024}. 
These methods, similar to the approach we propose,
are motivated by the notion
that variables with similar \red{outcome features} should have similar covariation patterns. However, these approaches, like the other popular factor model priors, shrink the covariance matrix towards a diagonal form.

We propose a method for high-dimensional covariance estimation
that utilizes auxiliary information on the outcome variables to inform a
covariance structure to shrink an unstructured population covariance matrix toward.
In particular, we detail a
prior specification that
decomposes an unstructured
covariance matrix as the sum of a
diagonal variance matrix and
a possibly low-rank covariance matrix that is shrunk towards 
\red{the matrix-square of the linear predictor of a matrix-variate regression based on the outcome features.}
We detail how the datatype of an outcome feature informs the centering of the induced covariance prior. Furthermore, this work uses a factor model representation to flexibly incorporate dimension reduction.
As such, our approach couples dimension reduction with structured shrinkage estimation. 
In contrast to other factor model priors that shrink towards diagonal covariance matrices, 
our approach utilizes the \red{outcome features} to shrink adaptively towards more flexible covariance structures. 

Related to our approach
is the recent work of
\cite{Heaps2024} that
proposes a framework for
shrinkage toward a structured covariance matrix in a latent factor model framework \red{via a non-exchangeable prior}.
\red{Specifically, the structure is incorporated via a mean-zero matrix-normal prior distribution.
While their proposed prior has mean-zero, 
the non-exchangeability allows for a non-sparse prior predictive covariance structure.}
This framework
requires user-specification of the structure, and parameter estimation procedures must be derived for each structure specification. 
In contrast, our framework
provides \red{a model-based method} to inform an appropriate structure from auxiliary information within one model, and correspondingly, one computational algorithm.
\red{Formally detailed in Section \ref{sec_cmr}, this is accomplished via a
non-zero-mean prior on the 
factor loading matrix, formulated as 
a multivariate regression with the outcome features.}
It is increasingly the case that such \red{outcome features} are available and easily accessible, so a framework that makes use of them in lieu of a
user-specification
is desirable.
\red{Related to our proposal, \cite{Pourahmadi1999} present a framework for integrating \red{outcome features in a regression} framework to estimate a precision matrix for longitudinal data.} 

The article proceeds as follows. 
In Section \ref{methods}, we elaborate on the limitations of existing methods for covariance estimation with TESIE data,
detail the proposed \red{feature aware covariance estimation} model to alleviate these shortcomings, 
and outline parameter estimation.
In Section \ref{CMR_implications}, we describe implications of the feature aware covariance estimation framework. 
In Section \ref{simulation}, we demonstrate the usefulness of the proposed approach in a simulation study and elaborate on its potential usefulness with data types typical in environmental health applications.
In Section \ref{application}, 
we analyze exposures from the TESIE study. In particular, we show improved inference on covariation in exposures and how the proposed modeling approach improves accuracy in imputing values below a limit of detection. We conclude with a discussion in Section \ref{discussion}.

\section{Data Description and Modeling}\label{methods}

\subsection{TESIE Study}



The data from the TESIE study contain measurements of exposures
to 21 SVOCs, each measured from two exposure assessment tools: household dust and
silicone wristbands worn for several days by young children in the household. 
In this dataset, there are $n = 73$ independent
samples with 100\% detection for all $p= 42$ measurements. These chemicals are sub-categorized into one of three
chemical classes. In particular, there are nine organophosphate esters (OPEs) \citep{Phillips2018}, commonly
flame retardants and plasticizers, five phenolic/paraben compounds \red{(phenols)} \citep{Levasseur2021}, found in plastics and personal
care products, and seven phthalates \citep{Hammel2019}, also found in plastics and personal care products.


Common approaches for analyzing multivariate exposure data assume that exposures for
chemicals in different classes are independent or alternatively collapse data prior to analysis to class-specific summaries \citep{Zhu2024}.
However, the classes are coarse-scale groups based on similarities in chemical structure, which is not necessarily the primary driver of correlation in the exposures data. 
In particular, chemicals belonging to different classes can have similar uses and thus exposure to these chemicals could be positively correlated.
For example, people with an increased exposure to phthalates found in certain personal care products may also have a higher exposure to parabens, \red{a subset of chemicals in the phenolic chemical class}, that tend to be found in similar personal care products.
However, chemical class information should not be discarded as class can be informative about covariance structure, and within-class inferences are of practical interest since class labels are used for product labeling and policy making.

Similar division of exposures may occur on the basis of the measurement assessment tool, such as urine, blood, dust, among others. 
In the TESIE data included here, exposures are measured from dust and silicone wristbands.
Measurements from the dust samples contain exposure information from a single micro-environment in the child's home, whereas the wristbands contain exposure information from multiple micro-environments over the course of several days. Analyzing data from both assessment tools jointly, and allowing for non-zero across-tool correlations, may allow for a more accurate summary of subjects' exposure profiles.

Ideally, 
all chemical exposures measured in a study would be analyzed jointly,
allowing information on chemical classes and exposure pathways to inform the covariance in a flexible manner.
Understanding covariance in exposure is of key interest in environmental epidemiology.
In particular, inferring the patterns of exposure among mixtures has been identified as a key research goal in the field
\citep{Joubert2022,Gibson2019,Zhu2024}.
Accurate summaries of exposure patterns through covariance estimation in a population can aid in the identification of vulnerable groups and targeted policy interventions. In addition, accurate inferences on covariance can improve handling of missing or censored exposure data, including issues with certain exposures measured below the limit of detection.

\begin{figure} 
\centering
\includegraphics[keepaspectratio,width=.49\textwidth]{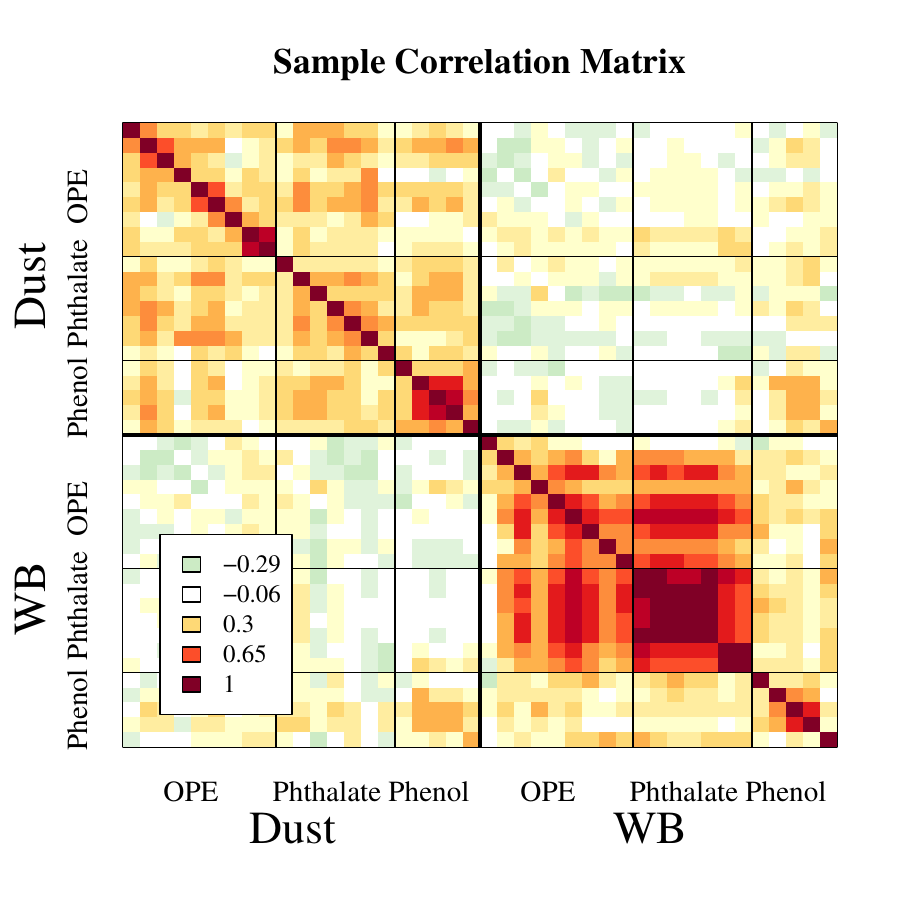}
\includegraphics[keepaspectratio,width=.49\textwidth]{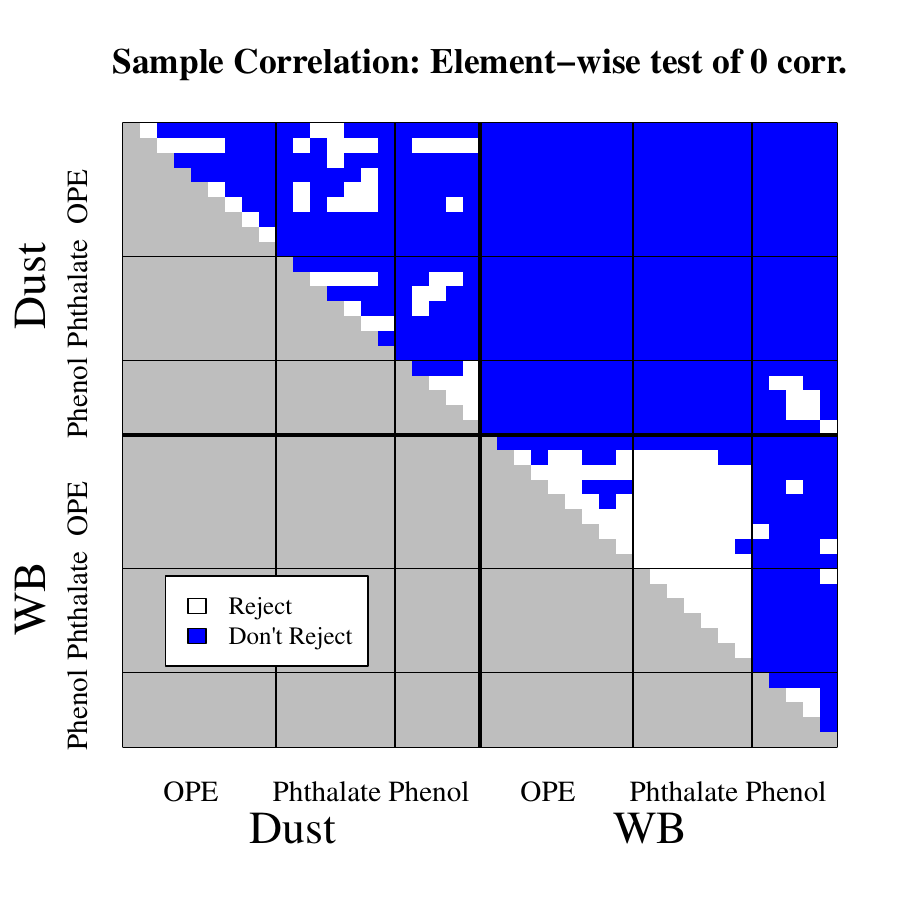}
\caption{Analysis of the TESIE data using a naive approach. Sample correlation (left) and failure to reject the null hypothesis of element-wise zero correlation with 5\% FDR control (right).
}\label{fig:MLE_TESIE}
\end{figure}

To motivate our approach, 
we first analyze the covariance among exposures from the TESIE data using standard approaches. 
In the left panel of Figure \ref{fig:MLE_TESIE},
the sample covariance matrix of the TESIE sample is plotted on the correlation scale for ease of visualization. The chemical compound abbreviations are: EHDPP, TCEP,TCIPP, TDCIPP,
TPHP, 2IPPDPP, 24DIPPDPP, 4tBPDPP, B4TBPP, DEP, DiBP, BBP, DBP, DEHP, DINP, TOTM, BPA, EPB, MPB, PPB, TCS. For more detailed descriptions of the chemicals, including full chemical names, see Appendix A.
Throughout this article, we plot all correlation matrices using the color palates \textit{GnBu} and \textit{YlOrRd},  from the R package \texttt{RColorBrewer} \citep{Neuwirth2014}, to represent negative (green) and positive (red) correlations, with the colors deepening in shade as values move away from zero. Zero values are represented by white blocks.
In the right panel of Figure \ref{fig:MLE_TESIE},
the results of the zero correlation hypothesis tests are plotted with 5\% false discovery rate (FDR) control, using the Benjamini-Yekutieli multiplicity adjustment \citep{Benjamini2001}. The blue blocks indicate a failure to reject the null hypothesis of zero correlation between the corresponding variables.

As evidenced by the sample covariance matrix, 
there is some shared \red{correlation structure among exposures, 
particularly among 
those measured from the same assessment tool.} 
For example, \red{the three parabans in the phenol chemical class, EPB,
MPB, and PPB,} 
exhibit a strong positive correlation 
\red{among measurements collected from dust samples and among measurements collected from wristbands}, 
and these relationships are statistically significant. 
This is expected as these parabens are commonly used in combination in personal care products, including lotion \citep{Guo2014}. In general, though,
the empirical covariance or correlation matrix has a high error. 
Few correlations are determined to be statistically significantly different from zero.
In particular, the null hypothesis of zero correlation is not rejected for 98.4\% of 
the pairs of 
\red{
cross-covariances between dust and wristband exposures, that is, in the off-diagonal block.}
Given the sample size constraint of these data, 
it seems likely that this is due to low power and not to the true correlation structure being so highly sparse.

\subsection{Covariance Modeling via a Latent Factor Representation}\label{sec_lfm}

For high-dimensional data such that the covariance dimension $p\times p$ is large relative to the number of samples $n$, a latent factor model that represents an unstructured covariance matrix as the sum of a diagonal variance matrix and a possibly low-rank covariance matrix is often utilized. This factor representation flexibly reduces the number of unknown parameters in the estimation of a covariance matrix so that it may be estimated with reduced error. 
Formally,
let $\ybf_1,\dots,\ybf_n$ be an independent and identically distributed (i.i.d.) random sample of $p$-dimensional
vectors from a mean-zero normal population with unknown non-singular covariance matrix $\Sigmabf\in\mathcal{S}_p^+$,
\begin{equation}\label{normaldist}
\ybf_1,\dots,\ybf_n\sim N_p(0,\Sigmabf).
\end{equation}
Data are typically centered prior to analysis so that the mean zero assumption is reasonable. 
Then, a factor model representation uses a parameter expanded framework defined as follows,
\begin{equation}\label{eqn:factormodel}
\ybf_i = \Lambdabf\etabf_i + \epsilonbf_i, \quad \epsilonbf_i\sim N_p(0,\Dbf),\quad\text{independently for } i =1,\dots,n, 
\end{equation}
where 
$\Lambdabf$ is a ($p \times r$)-dimensional real-valued matrix, often referred to as a factor loading matrix,
${\etabf}_1,\dots,\etabf_n\sim N_r(0,\eye_r)$ are i.i.d. latent factors, and \red{$\epsilon_1,\dots,\epsilon_n$ are idiosyncratic noise} with diagonal covariance
$\Dbf$, 
\red{whose entries $d_1,\dots,d_p$ each represent unique variance not attributable to the factors} \citep{Mardia1979}.
Marginal with respect to the latent factors, the response variables follow a multivariate normal distribution with covariance matrix $\Sigmabf$ where
\begin{equation}\label{latentfactcov}
\Sigmabf = \Dbf + \Lambdabf\Lambdabf^T.
\end{equation}

There are a few primary benefits to using such a decomposition of the population covariance matrix. 
For one, representation (\ref{latentfactcov}) consists of 
$p (r+1)$ unknown parameters, which, depending on the choice of $r$, can be markedly less than that of an unstructured, unconstrained covariance matrix
consisting of $p (p+1)/2$ unknown parameters. 
Another benefit of the latent factor covariance representation
is that the entries of the factor loading matrix may be any real value and the resulting covariance matrix $\Sigmabf$ will be positive definite. 
In what follows, we show that this presents an avenue to utilize a prior that makes use of auxiliary knowledge regarding the structure of the covariance matrix that can enable straightforward computation without cumbersome computational constraints to
enforce positive definiteness.
For a discussion of additional benefits of a latent factor representation, see \cite{Fruhwirth2025}.

To this end, \red{the sharpness of the covariance estimates in the TESIE data may be improved by using a latent factor model}.
When \red{outcome features} are available, one possibility is to utilize a latent factor model with the structured shrinkage prior of \cite[SIS]{Schiavon2022}. 
\red{Specifically, the SIS model uses a mean-zero normal prior on the factor loading parameters with variance that is a product of three independent variables: a global variance, a factor-specific variance that induces increasing shrinkage as the factor index increases, and a local variance unique to each of the factor loading elements. 
This local variance invokes non-exchangeability and is modeled such that the expected value is a transformed regression on outcome features.}
In this way, SIS uses \red{outcome features} to inform the sparsity structure in the covariance matrix. 
\red{Furthermore, the SIS model allows for infinitely many factors, 
but increasingly shrinks the columns of the factor loading matrix to zero  
via the factor-specific variance terms}.

\begin{figure} 
\centering
\includegraphics[width=.49\textwidth]{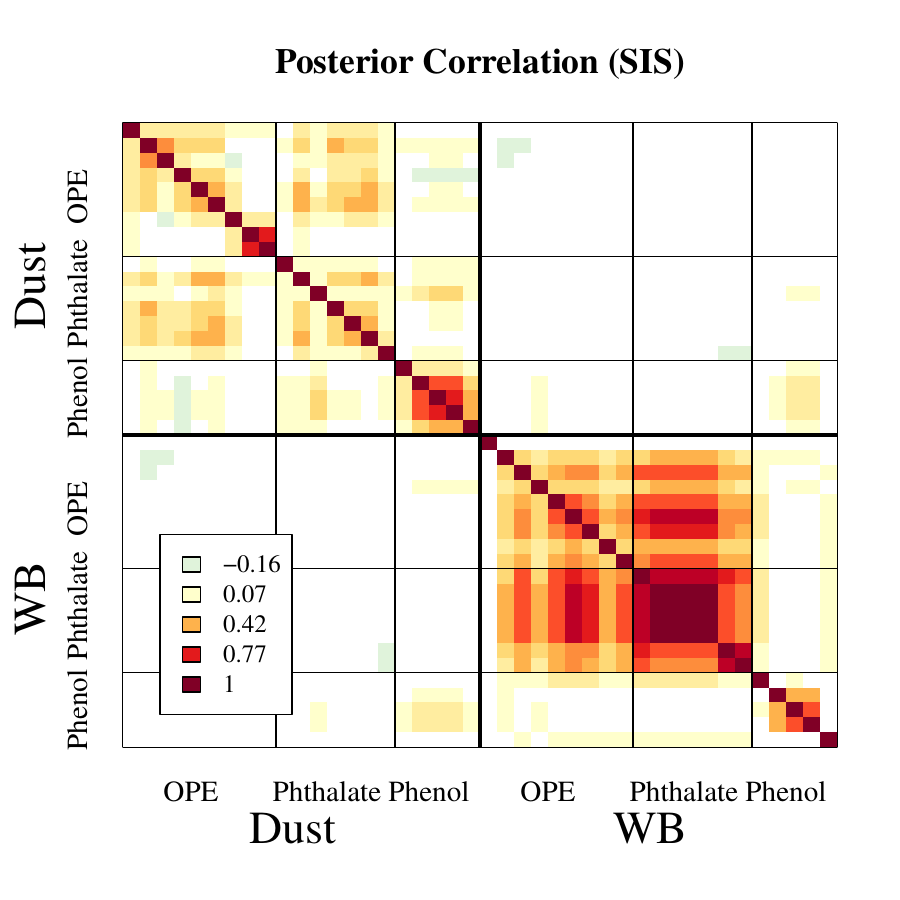} 
\includegraphics[keepaspectratio,width=.49\textwidth]{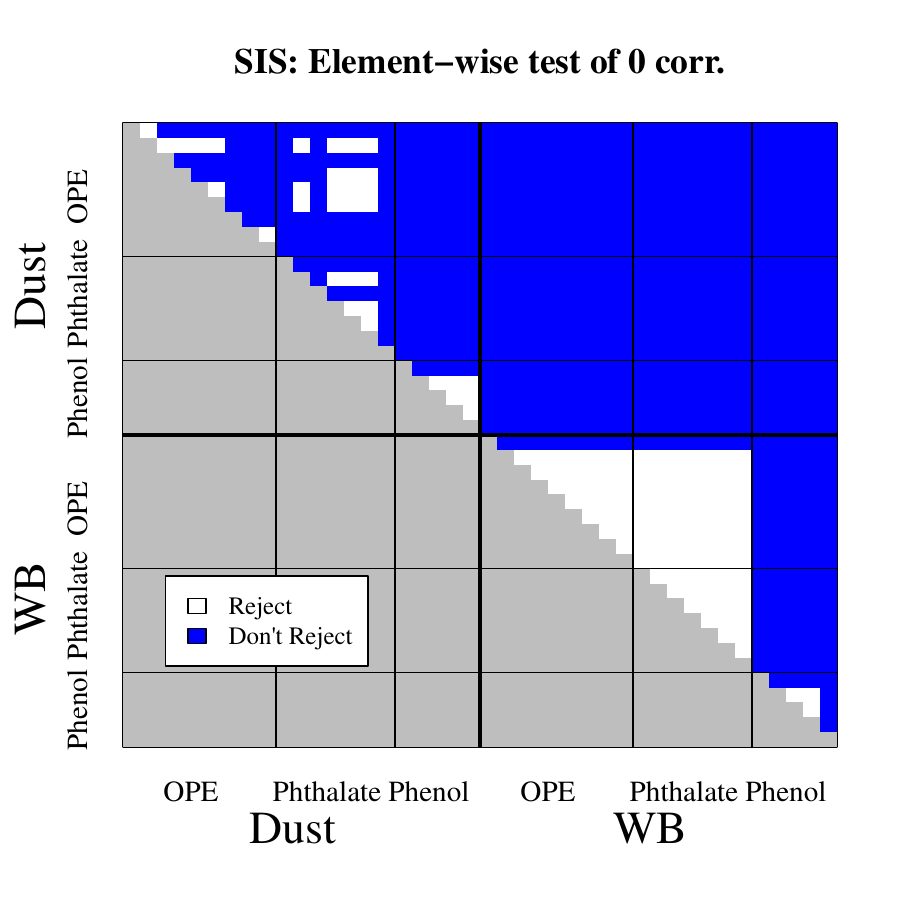}
\caption{Analysis of the TESIE data using the SIS method.  Posterior correlation (left) and inclusion of 0 in an element-wise 95\% credible interval (right). }\label{fig:SIS_TESIE}
\end{figure}

\red{We analyze the TESIE data with the SIS model using code provided by the authors \citep{Schiavon2022a} with the following outcome features (discussed in detail in Section \ref{application}): Chemical class membership, exposure assessment tool, chemical name, vapor pressure, and high production volume indicator.}
The Bayes estimator of the covariance matrix is plotted on a correlation scale in the left panel of Figure \ref{fig:SIS_TESIE}. The SIS estimate shrinks many of the correlations closer to zero compared to the MLE. Due to shrinkage, element-wise 95\% credible intervals are significantly narrower than 95\% bootstrapped confidence intervals on the MLE. 
This reflects a potential improvement in efficiency, 
however, because the shrinkage is towards zero,
most element-wise 95\% SIS credible intervals contain zero (right panel of Figure \ref{fig:SIS_TESIE}). 
These results motivate analyzing the TESIE data with an approach that allows for shrinkage towards non-\red{diagonal} correlation structures in the data informed by the \red{outcome features}.

\subsection{Feature Aware Covariance Estimation Prior}\label{sec_cmr}


Prior distributions on the factor loading matrix typically favor shrinkage towards zero, often such that the prior mean of the loadings matrix is the zero matrix. By the factor decomposition of the covariance matrix (Equation \ref{latentfactcov}), a zero-factor loading matrix corresponds to a diagonal prior predictive covariance matrix. In this way, priors on the factor loading that shrink elements towards zero tend to shrink the induced population covariance towards a diagonal structure.
Such priors can shrink away interesting structures in high-dimensional covariance matrices. 

We seek to improve covariance estimation accuracy by shrinking towards a more flexible structure informed by auxiliary information provided via \red{outcome features.}
To this end, 
we propose the \textit{feature aware covariance estimation} (FACE) prior,
defined such
that row $j\in\{1,\dots,p\}$ of the factor loading matrix, corresponding to outcome variable $j$, is a $r$-dimensional regression on the corresponding \red{outcome features}. That is, for variable $j$,
\begin{equation}\label{CMRj}
\lambdabf_j
\sim N_r( 
\Gammabf^T\xbf_j,d_j\Tbf ),\quad\text{independently for } j =1,\dots,p,
\end{equation}
where $\xbf_j$ is a $q$-dimensional vector of meta covariates (indexed by variables instead of samples),
$\Gammabf\in\reals^{q\times r}$ is a regression coefficient matrix, \red{$d_j$ represents the idiosyncratic variance for variable $j$ (Equation \ref{eqn:factormodel}),}
and $\Tbf \in\mathcal{S}_r^+$ is a $r$-dimensional covariance matrix \red{representing the covariance among the $r$ factors}.
Equivalently, the FACE prior may be expressed in matrix form,
\begin{equation}\label{CMR}
\Lambdabf \sim N_{p\times r}(
\Xbf\Gammabf,\Tbf \kron \Dbf),
\end{equation}
where $\;\kron\;$ denotes a Kronecker product,
and $\Xbf$ is a $(p\times q)$-dimensional \red{outcome feature} matrix for all $p$ response variables. 
\red{Here, again, $\mathbf{\Gamma}$ is the outcome feature regression coefficient such that the prior predictive estimator of the factor loading matrix is $\mathbf{X}\mathbf{\Gamma}$. This coefficient in turn determines the prior predictive covariance matrix.
}

\red{Marginalizing out the factor loading matrix $\Lambdabf$ from the factor model, Equation \ref{eqn:factormodel}, results in an implied regression of each response $y_i,\;i\in\{1,\dots,n\}$ on a constant covariate $\Xbf$ with non-normal errors and regression coefficients unique to each sample unit. Specifically, marginal with respect to $\Lambdabf$, the factor model can be expressed as the following,
\begin{equation}
    \ybf_i= \Xbf\tilde{\Gammabf}_i +\tilde{\epsilon}_i,
\end{equation}
where $\tilde{\Gammabf}_i=\Gammabf\eta_i$ and $\tilde{\epsilon}_i=\epsilon+\varphi\eta_i, \;\varphi\sim N_{p\times r}(0,\Tbf\kron\Dbf)$.
}
In Section \ref{CMR_implications}, we elaborate further on the implications of the FACE prior, and provide details for various data types of covariate information commonly seen in practice.

\subsection{Parameter Estimation}\label{CMR_estimation}

Bayesian posterior computation under the FACE model is straightforward with
closed-form full conditionals of the covariance parameters, $\Lambdabf,\Dbf$. Inference proceeds based on the joint posterior distribution with density
$p(\Dbf,\Lambdabf,\Tbf,\Gammabf|\ybf_1,\dots,\ybf_n)$.
For a full Bayesian analysis, priors must be specified for all unknown parameters.
\red{In this work, we set $\mathbf{T}=\tau^2\mathbf{\Theta}$} for diagonal $\Thetabf$ \red{to enable flexible shrinkage of the low-rank dimension $r$.}
First, we suggest weakly informative semi-conjugate priors
on the variances $\{d_j\}_{j=1}^p$, FACE regression coefficient matrix $\Gammabf$, and FACE prior \red{global} scale $\tau^2$,
\begin{align}
    \Gammabf \sim{}& N_{q\times r} (\boldsymbol{0},\Thetabf \kron \eye_q),\nonumber\\
    \{d_j\}_{j=1}^p \sim{}& InverseGamma(a_d/2,b_d/2),\\
    \tau^2\sim{}& InverseGamma(a_\tau/2,b_\tau/2).\nonumber
\end{align}

The prior specification of the variance matrix $\Thetabf$ requires particular attention.
\red{To enable shrinkage of the low rank dimension $r$, we place an increasing shrinkage prior on the elements of the diagonal covariance matrix $\mathbf{\Theta}$ such that the elements are increasingly concentrated around zero via the prior proposed in \cite[CUSP]{Legramanti2020}.}
$\Thetabf$ affects the posterior concentration of both the 
\red{outcome feature} regression coefficient matrix and the resulting factor loading matrix. If an element of $\Thetabf$ is approximately zero, the corresponding columns of the coefficient and factor loadings will be concentrated around zero. 
\red{Specifically, for $h=1,\dots,r$ indexing the diagonal elements of $\Thetabf$,}
\begin{equation}\label{eqn:cusp_specification}
\theta_h|\pi_h \sim (1-\pi_h)IG(a_\theta,b_\theta)+\pi_h\delta_{\theta_\infty} ; \quad \pi_h=\sum_{l=1}^h\omega_l;\quad \omega_l=\nu_l\prod_{m=1}^{l-1}(1-\nu_m),
\end{equation}
and $\nu_1,\dots,\nu_r\sim Beta(1,\alpha)$.
\red{Here, $\delta_{\theta_{\infty}}$ is a spike at a near-zero value $\theta_\infty$, 
to facilitate shrinkage of redundant factors.}
The posterior computation for the CUSP parameters proceeds with a data augmentation scheme that introduces $z_h\sim Categorical(\boldsymbol{\pi})$.

\red{Because we are using a cumulative shrinkage prior on the number of factors $r$, $r$ can be considered an upper truncation of the possible number of active factors.
To this end, $r$ should be set to be large enough to characterize the true population covariance, and \cite[\S 3.3]{Legramanti2020} show $(p+1)$ suffices as a conservative maximum. 
To mitigate this conservatism, $r$
may be adaptively tuned in the Gibbs sampler, as proposed in \cite{Bhattacharya2011}, however, we do not use adaptive tuning in our implementations.}

For this prior specification, the unknown parameters in the FACE model maintain semi-conjugacy leading to a straightforward Gibbs sampler algorithm that constructs a Markov chain in all unknown model parameters. The resulting posterior samples provide a basis for Bayesian inferences on any functional of the unknown parameters, including point estimation and uncertainty quantification. The Gibbs sampler proceeds by iteratively sampling the model parameters from their full conditional distributions until convergence. The resulting algorithm produces a Markov chain
with stationary distribution corresponding to the joint posterior distribution of interest. The sampling steps are detailed in Algorithm \ref{CMRalgorithm}.

\begin{algorithm}\caption{One iteration of the FACE Gibbs sampler}\label{CMRalgorithm}
    \begin{enumerate}
    \item For $i=1,\dots,n$: Sample 
    $\etabf_i$ from $N_r(\Sbf_\eta^{-1}\Lambdabf^T\Dbf^{-1}\ybf_i,\Sbf_\eta^{-1})$ where $\Sbf_\eta = \Lambdabf^T\Dbf^{-1}\Lambdabf+\eye_k$.
    \item For $j=1,\dots,p$: Sample $d_j$ from $IG( (n+r+a_d)/2,S_{dj}/2)$
    where
    \[
    S_{dj} = \sum_{i=1}^n (y_{ij}-\lambdabf_j^T\etabf_i)^2 + 
    ||\lambdabf_j -\Gammabf^T\xbf_j||^2_{\Tbf^{-1}} +b_d.
    \]
    \item Sample $\Lambdabf$ from $N_{p\times r}([\Xbf\Gammabf     \Thetabf^{-1}/\tau^2+\sum_{i=1}^n\ybf_i\etabf_i^T]\Sbf_\Lambda^{-1}, \Sbf_\Lambda^{-1}\kron \Dbf)$
    where
    \[
    \Sbf_\Lambda =\Thetabf^{-1}/\tau^2+\sum_{i=1}^n\etabf_i\etabf_i^T.
    \]
    \item Sample $\Gammabf$ from $N_{q\times r}(\Sbf_\Gamma^{-1} \Xbf^T\Dbf^{-1}/\tau^2\Lambdabf,\Thetabf \kron \Sbf_\Gamma^{-1})$ where $\Sbf_\Gamma = \Xbf^T\Dbf^{-1}/\tau^2\Xbf +\eye_q $.
    \item Sample 
    $\tau^2$ from $IG( (pr+a_\tau)/2 , S_\tau/2)$ where
    $
    S_\tau = tr(||\Lambdabf-\Xbf\Gammabf||^2_{\Dbf^{-1};\Thetabf^{-1}}) + b_\tau
    $.
    \item Sample $\Thetabf$ from a modified CUSP Gibbs sampling procedure: 
    \begin{enumerate}
        \item For $h=1,\dots,r$: Sample $z_h$ from a categorical distribution with probabilities based on,
        \[
        pr(z_h=l|\cdot) \propto \begin{cases}
            \omega_l N_p(\lambda_j;0,\theta_\infty \eye_p), &l=1,\dots,h\\
            \omega_lt_{2a_\theta}(\lambda_j;0,(b_\theta/a_\theta)[\Xbf\Xbf^T+\tau^2\Dbf])&l=h+1,\dots,r.
        \end{cases}
        \]
        \item For $l=1,\dots,r$: Sample $\nu_l$ from $Beta(1+\sum_{h=1}^r\one(z_h=l),\alpha + \sum_{h=1}^r\one(z_h>l))$,
        and set $\nu_r=1$.
        \item For $l=1,\dots,r$: Compute
        $\omega_l= \nu_l\prod_{m=1}^{l-1}(1-\nu_m)$.
        \item For $h=1,\dots,r$: If $z_h\leq h,$ set $\theta_h=\theta_\infty$; otherwise, sample $\theta_h$ from 
        \[
        IG(a_\theta + p/2,b_\theta + (\lambda_h^T[\Xbf\Xbf^T+\tau^2\Dbf]^{-1}\lambda_h)/{2})
        \]
    \end{enumerate}
\end{enumerate}
\end{algorithm}

\section{Structural Implications of Feature Aware Covariance Estimation}\label{CMR_implications}

In this section, we highlight the implications of the FACE model on covariance estimation
by detailing the form of the \red{prior predictive
covariance matrix, that is, the covariance of the marginal distribution of $\ybf$}. 
For a general \red{feature outcome} matrix $\Xbf$,
it can be shown that the prior \red{predictive} covariance is 
the sum of a diagonal variance matrix and a covariance matrix with rank $\min(q,r)$ defined by the \red{feature outcome} regression:
\begin{equation}\label{genCMRReg}
    Cov(\ybf|\red{\Gammabf,\Tbf})= \tilde{\tau}^2\eye_p  +\Xbf\Gammabf\Gammabf^T\Xbf^T,
\end{equation}
where $\tilde{\tau}^2 = (1+tr(\Tbf))\red{b_d/(a_d-2)}$. See Appendix B for derivation of Equation \ref{genCMRReg}.
\red{
In the discussion that follows, for notational convenience, all statements are understood to be conditional on 
$\Gammabf,\Tbf$ unless explicitly noted otherwise.}
For different data types of the \red{feature outcomes}, the form of this prior \red{predictive} covariance matrix reduces to various familiar structures.
In what follows, we elaborate on this phenomenon for a few specific data types of \red{feature outcomes} that are often available in practice.
\red{We note that computation for the FACE model remains the same regardless of the datatype of the outcome features.}

\subsection{Intercept Model}\label{sec:FACE_I}

As a baseline regime, we consider a case that does not utilize any relevant auxiliary data for the variables. In this context, 
an intercept, $x_1=\dots=x_p=1$,
may be used as a default covariate,
and the prior \red{predictive} covariance matrix simplifies to a \textbf{compound symmetric} matrix, 
\begin{equation}\label{eqn:cmr_intercept}
Cov(\ybf)= \tilde{\tau}^2\eye_p  + \gamma^*\one_{p\times p},
\end{equation}
where $\gamma^* =\Gammabf \Gammabf^T$ for $\Gammabf\in\reals^{1\times r}$.
That is, for covariance estimation with the FACE prior using $\Xbf = \one_p$, the covariance matrix is shrunk towards a compound symmetric matrix where the off-diagonal term is determined flexibly by the cross product of a $r$-dimensional vector. 

\subsection{Categorical Model} \label{categoricalmodel}

An \red{outcome feature} that assigns one of $q<p$ labels to the $p$ variables is 
a common instance of auxiliary information that may be used in the FACE framework.
In some applications,
such labels will correspond to a grouping of the variables that
is used to motivate an assumption of zero correlation across groups, and variables belonging to different groups may be analyzed separately; 
such a restrictive framework implicitly imposes a block-diagonal covariance structure on the population covariance matrix.
Other statistical methodologies
motivated by the use of group information of variables include group lasso regression \citep{Yuan2006} and multilevel methods \citep{Gelman2007,Rao2015}.

However, in many applications, the assumption of zero correlation between groups is inappropriate. Instead, grouping information may be used in the FACE framework to inform the structure toward which a covariance estimate is shrunk. Specifically, to incorporate a categorical grouping of the variables into the FACE prior, it may be encoded as a categorical design matrix $\Xbf$ that is formed based on group membership. Formally, let
$\xbf_j=s(g_j)\in\{0,1\}^q$ 
be an indicator variable 
denoting the group membership of variable $j$ in a categorical 
meta covariate $\textbf g \in \{1,\dots,q\}^p$. Then, for variable $j\in\{1,\dots,p\}$, the FACE prior on row $j$ of the factor loading matrix (Equation \ref{CMRj}) simplifies to a multivariate random intercept model,
\begin{equation}
\lambdabf_{j} \sim N_r(\gammabf_{g_j},d_j\Tbf),
\end{equation}
where $\gammabf_{g_j}$ is 
defined to be the $g_j$th row of $\Gammabf$, that is, the $r$-dimensional regression coefficient for group $g_j$. In this context, the prior \red{predictive} covariance matrix induced by the FACE prior is a \textbf{block covariance matrix} defined such that the off-diagonal covariances are determined by an inner product of the corresponding random intercepts:
\begin{equation}\label{eqn:prior_marg_cov_cat}
    Cov(y_{j},y_{j'}) = \tilde{\tau}^2+ \gammabf_{g_j}^T\gammabf_{g_{j'}},\quad j,j'\in\{1,\dots,p\}. 
\end{equation}
\red{Of note, when $g_j=g_{j'}$, the prior \red{predictive} covariance specified in Equation \ref{eqn:prior_marg_cov_cat} will be positive. That is, the prior correlation among elements with shared group membership will be positive.}

\subsection{Multiple Categorical Model}\label{multiple_cat_case}  

Closely related to using a single categorical class label as an \red{outcome feature} is a case where there is information regarding membership of multiple categorical classes for each of the response variables. For the chemical exposure data collected in TESIE, chemicals each belong to one of three chemical classes and have been measured from one of a few exposure pathways. Other application areas where multiple class labels are available for each variable
include text or image classification \citep{Papadopoulos2022}, species modeling \citep{Stolf2024}, and spatial modeling in a gridded domain \citep{Peruzzi2022}.

Defining appropriate notation, $I$ categorical variables each comprised of $q_i$ categories for $i\in\{1,\dots,I\}$ can be encoded in a design matrix $\Xbf$ and incorporated into the FACE framework. Let $\textbf{g}^{(i)}\in\{1,\dots,q_i\}^p$ be the $i$th categorical \red{outcome feature}, and define $q = \sum_{i=1}^I q_i$. Then, define the \red{outcome covariate} for variable $j\in\{1,\dots,p\}$ as the concatenation of the corresponding group membership of each categorical variable,
\begin{equation}
{\xbf_j} = \begin{bmatrix} s\big(g^{(1)}_j\big)^T &
\hdots &
s\big(g^{(I)}_j\big)^T\end{bmatrix}^T\in\{0,1\}^q.
\end{equation}
Then, for such an \red{outcome covariate}, 
the FACE prior on row $j$ of the factor loading matrix simplifies to an additive sum of group mean effects,
\begin{equation}\label{multiple_categorical_cmr}
\lambdabf_j \sim N_r\Big(\sum_{l = 1}^q \gammabf_l \indicator_{(x_{jl} = 1)} , d_j\Tbf\Big),
\end{equation}
where $\gammabf_l$ is defined to be the $l$th row of $\Gammabf$.
The corresponding prior \red{predictive} covariance is 
\begin{align}\label{eqn:prior_marg_cov_mult_cat}
    Cov(y_{j},y_{j'}) ={}& \tilde{\tau}^2 + E(\lambdabf_j)^TE(\lambdabf_{j'}) \\
    ={}& \tilde{\tau}^2 +\Big(\sum_{l = 1}^q \gammabf_l \indicator_{(x_{jl} = 1)}\Big)^T
    \Big(\sum_{l = 1}^q \gammabf_l \indicator_{(x_{j'l} = 1)}\Big),\quad j,j'\in\{1,\dots,p\}.\nonumber
\end{align}

As an illustrative example of the impact of the FACE prior in this context, consider two toy \red{outcome feature} vectors, $\xbf^T_j = \begin{bmatrix} 1 & 0 & 1 & \textbf{0}^T\end{bmatrix}$ and $\xbf^T_{j'} = \begin{bmatrix} 1 & 1 & 0 & \textbf{0}^T\end{bmatrix}$. Then, the covariance between variables $j,j'$ based on these \red{outcome features} is determined by the cross product of prior marginal expectations of row $j,j'$ in the factor loading matrix,
\begin{equation}
E(\lambdabf_j)^TE(\lambdabf_{j'}) = 
( \gammabf_1 + \gammabf_3)^T
(\gammabf_1 + \gammabf_2 ). \nonumber
\end{equation}
In this way, the FACE prior 
in a regime that encodes multiple categorical class membership as \red{outcome features} corresponds to a flexible prior defined by additive and multiplicative interactive effects dependent upon the various variables' group membership profile.

\subsubsection{Modeling Matrix-variate Data as a Multiple Categorical Model}\label{sec:cmr_matrixvariate}

The multiple categorical FACE model can account for the inherent matrix structure present in many exposure mixture datasets. Such data often consist of measurements of exposures to a set of chemicals from multiple sources of exposure through different pathways \citep{Dixon2018,Hammel2016}. 
Such repeated measurements of a multivariate response can be expressed as a matrix-valued data point with $p_1$ rows corresponding to exposure pathways and $p_2$ columns corresponding to chemicals.
In practice, an analysis of such matrix-variate data will commonly impose a separable covariance structure such that the population covariance is represented as the Kronecker product of a
$p_1$-dimensional `row' covariance matrix and a $p_2$ dimensional `column' covariance matrix
\citep{Dawid1981}. 
The matrix-variate structure of the TESIE data is exploited in an analysis of subgroups in the data in \cite{Bersson2024a}.
In addition to the matrix-structure of the variables in the TESIE data, there are a wide array of additional auxiliary information which we aim to exploit in this work.

In general, the suitability of a Kronecker structural assumption is often unclear, as discussed, for example, in \cite{Stein2005}.
To this end, recent work develops models that are robust to the separable covariance assumption
\citep{Hoff2023b,Masak2023}.
Estimates from these analyses may be unstable, though, if the population is not well represented by a separable covariance matrix.
Alternatively, we propose modeling the covariance matrix for such data with the FACE model where the matrix-variate structure is encoded as multiple categorical meta
covariates that indicate
row and column membership of each variable. 
In this way, matrix-variate data may be analyzed as a special case of the multiple categorical meta covariate model.
What is more, with a variable selection mechanism on the meta covariates, 
our approach can adapt more dynamically to the relevance of the matrix structure of the data in covariance estimation. 
For example, the FACE framework can adapt to the case where row membership is informative but the column membership is not informative in modeling the covariance matrix.

To construct \red{outcome features} that identify the row and column membership of variables for a matrix-variate data point $\Ybf\in\reals^{p_1\times p_2}$,
let $j\in\{1,\dots,p_1p_2\}$ index the variables in $\ybf:=vec(\Ybf)\in\reals^{p_1p_2}$,
where the vectorization operator $vec(\cdot)$ stacks the columns of the matrix taken as an argument.
Then, the \red{outcome feature} corresponding to variable $j$ is a $(p_1+p_2)$-dimensional vector identifying the row and column membership of variable $j$, defined as
\begin{equation}
{\xbf_j} = \begin{bmatrix} s(g^{(1)}_j) &
s(g^{(2)}_j)\end{bmatrix} \nonumber
\end{equation}
where $g^{(1)}\in \{1,\dots,p_1\}^{p_1p_2}$ denotes the row membership of $\ybf$ and $g^{(2)}\in \{1,\dots,p_2\}^{p_1p_2}$ denotes the column membership.
Then, the 
FACE prior for row $j$ of the factor loading matrix simplifies to the form given in Equation \ref{multiple_categorical_cmr}, and
correspondingly, the prior \red{predictive} covariance consists of additive and multiplicative row and column interactive effects. Specifically, for measurement $j$ corresponding to exposure to chemical $k$ from exposure pathway $l$ 
and measurement $j'$ of exposure to chemical $k'$ from pathway $l'$, the prior \red{predictive} covariance is
\begin{equation}\label{sep_cmr_cov}
    Cov(y_j,y_{j'}) = \tilde{\tau}^2 +(\gammabf_l +\gammabf_{p_1+k})^T(\gammabf_{l'} +\gammabf_{p_1+k'}).
\end{equation}
This approach allows for flexibility in covariance estimation based on matrix-variate data by incorporating the inherent matrix structure of the data without imposing a restrictive separable structural assumption.

\subsection{General Model}

In the most general case, available meta covariates may consist of continuous or mixed-data types.
As discussed, for example,
there are ample data on properties of chemicals commonly studied in environmental health applications including
categorical data such as chemical class and exposure pathway,
continuous data such as molecular weight and boiling point,
and ordinal data such as water absorption and environmental impact, 
to name a few.
Incorporation of generic \red{outcome features} 
in the FACE framework
allows for classical regression-type behavior in the context of covariance estimation.
Specifically, variables with similar \red{outcome features} will exhibit similar patterns of co-variation with the other variables. 
The more the \red{outcome features} deviate from one another, the more
flexibility the FACE prior affords with respect to structure of the prior \red{predictive} covariance matrix.

\section{Simulation Study}\label{simulation}

\begin{figure}
\centering
\includegraphics[keepaspectratio,width=.8\textwidth]{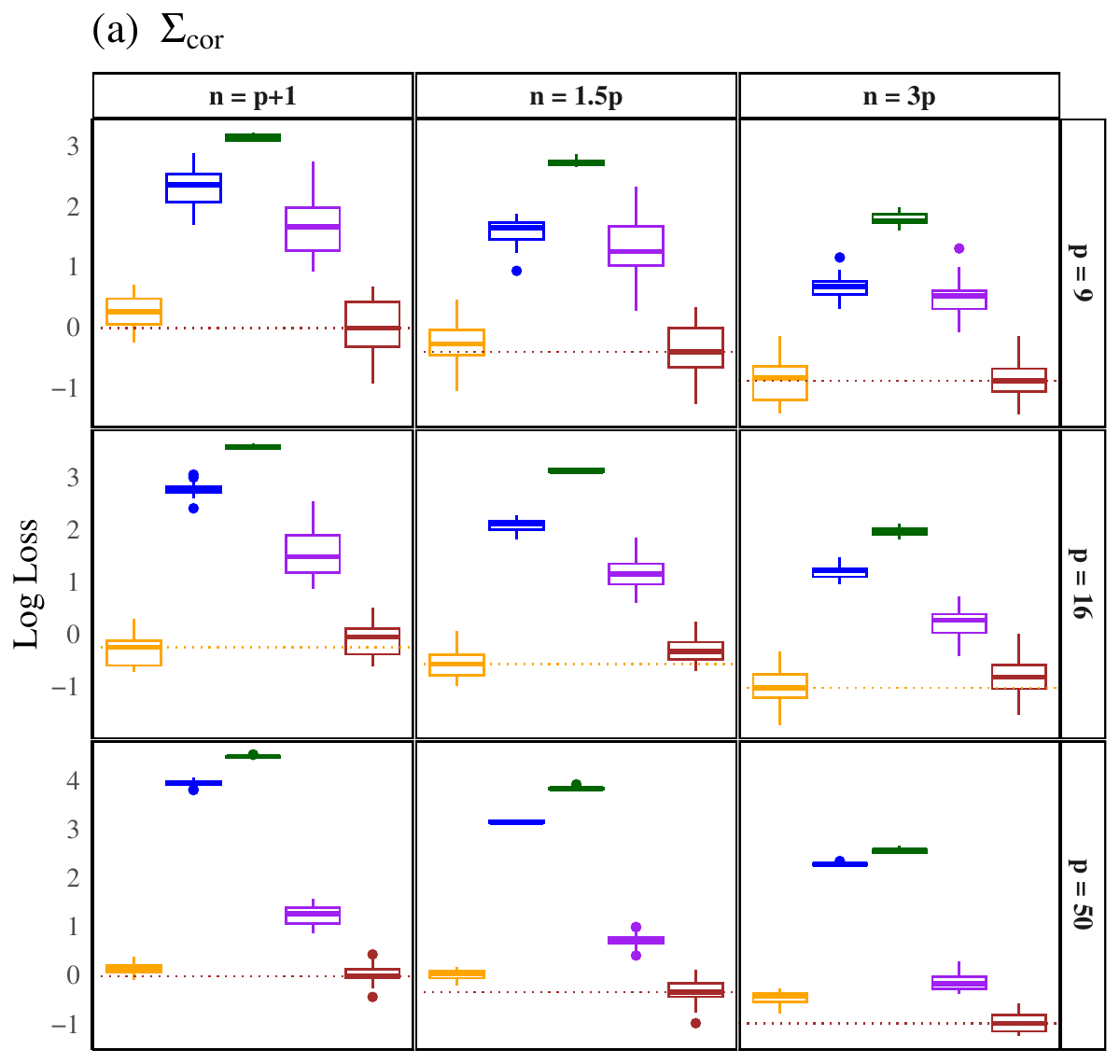}
\includegraphics[keepaspectratio,width=.8\textwidth]{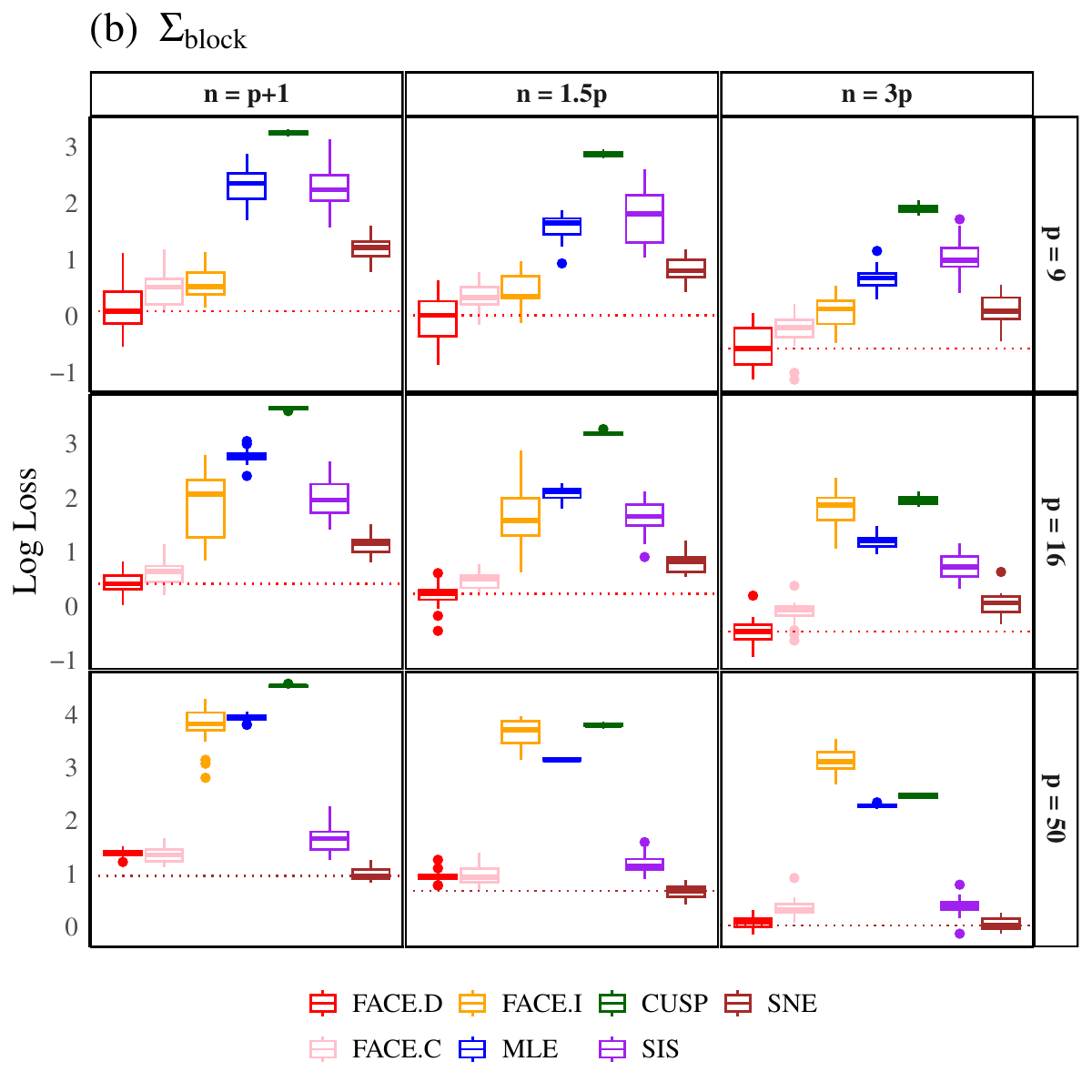}
\caption{
Boxplots of log Stein's loss for each regime, problem dimension, and sample size permutation. 
\red{Results plotted for FACE with an intercept (FACE.I),
FACE with group-identifying design matrix (FACE.D),
FACE with a continuous outcome feature (FACE.C),
MLE, 
cumulative shrinkage model (CUSP), 
structured shrinkage prior with group-identifying design matrix (SIS), and structured matrix normal prior (SNE). The  dashed lines denote the minimum median loss.}}\label{sim_study_main}
\end{figure}

In this section, we convey the usefulness of the proposed feature aware covariance regression prior through a loss comparison of covariance estimates obtained from popular priors proposed in the recent factor model literature. Motivated by inferring covariation among mixtures of exposures, we are particularly interested in accurately estimating the population covariance matrix in high-dimensional settings where the true population covariance matrix is \red{non-diagonal}.
To this end, we examine a regime where the population covariance matrix is 
an exchangeable correlation matrix with correlation 0.9 ($\Sigmabf_{cor}$) and
a regime with a blocked covariance matrix consisting of three groups with non-zero correlation among most groups ($\Sigmabf_{block}$).
\red{In the blocked regime, the correlation within groups 1, 2, and 3 is 0.75, .5, and .944, respectively. The correlation between group 1 and group 2 is 0.354, the correlation between group 2 and group 3 is 0, and the correlation between group 1 and group 3 is 0.118.}
In these settings, we demonstrate that utilizing the FACE prior, which encourages shrinkage towards a non-diagonal covariance matrix, improves accuracy over the MLE and commonly used priors for factor models.
Additionally, this study provides insight into the performance of the FACE model in both the presence and absence of various types of \red{outcome features} that inform the structure of the covariance matrix.

For each population covariance matrix regime, we consider dimensions $p\in\{9,16,50\}$ and sample sizes 
$n\in\{p+1,1.5p, 3 p\}$.
Then, for each covariance, dimension, and sample size permutation, we simulate $25$ datasets from a mean-zero normal population. 
For each dataset, we run the proposed Gibbs sampling algorithm for the intercept FACE model  (FACE.I) from \S \ref{sec:FACE_I}.
Additionally, for the block covariance regime, we run the Gibbs sampling algorithm for the FACE model with a group-identifying design matrix for the \red{outcome feature}
(FACE.D).
To represent a more realistic \red{outcome feature} state, we also compute estimates from the FACE model with a continuous \red{outcome feature} drawn from a normal distribution concentrated around a different mean for each group (FACE.C). 
Specifically, the \red{outcome feature} for each variable in the FACE.C regime includes an intercept and a real valued scalar obtained randomly from a normal distribution centered at the group number $g_j \in \{1, 2, 3\}$ with standard deviation 1/4. For example, the \red{outcome feature} matrix used for the smallest dimension $p=9$ is as follows, rounded to 2 decimal points:
\begin{equation}
\Xbf^T = \begin{bmatrix}
1 & 1 & 1 & 1 & 1 & 1 & 1 & 1 & 1 \\ 
0.86 & 0.94 & 2.39 & 2.02 & 2.03 & 3.43 & 3.12 & 2.68 & 2.83 \\ 
\end{bmatrix},
\end{equation}
where $\mathbf{g}^T = \begin{bmatrix} 1&1&2&2&2&3&3&3&3\end{bmatrix}$.
This \red{outcome feature} configuration is motivated by the intuition that, in practice, units belonging to a group have an affinity, so are expected to have similar \red{outcome feature} values.
\red{In all FACE implementations, we use weakly informative hyperpriors, and, specifically for the unknown parameters for the increasing shrinkage prior specification for the latent dimension, we set $r=\lceil (p-1)/2 \rceil$, $(a_\theta,b_\theta) = (1/2,1/2)$, $\theta_\infty = 0.05$, and $\alpha = \lfloor (p/3) \rfloor$.}

Estimates obtained from the FACE model are compared with the output of
the cumulative shrinkage model proposed in \cite{Legramanti2020} (CUSP).
\red{CUSP applies the prior specified in Equation \ref{eqn:cusp_specification} on the variance of a mean-zero normal distribution for a given factor loading with factor index $h=1,\dots,r$, for variable $j=1,\dots,p$, $\lambda_{jh}\sim N(0,\theta_h)$.
CUSP induces increased shrinkage towards zero of the factor loading as index $h$ increases.}
We also compare to the structured increasing shrinkage model \citep[SIS]{Schiavon2022} discussed in Section \ref{sec_lfm}.
For the correlated regime, SIS is run with no \red{outcome features}, and, for the block covariance regime, SIS is run with the same group-identifying \red{outcome feature} design matrix used in the FACE.D model.

\red{Additionally, we compare with the structured matrix normal prior for an exchangeable matrix from \cite{Heaps2024} (SNE).
SNE places a prior on the factor loading matrix such that the 
prior predictive covariance matrix is
an exchangeable correlation matrix. This structure is incorporated via dependent priors with mean-zero on the factor loading vectors. 
This is in contrast to the FACE prior for factor loadings vectors that have non-zero mean.}

All MCMC samplers are run for 20,000 iterations removing the first 10,000 iterations as a burn-in period. With the Markov chains obtained from each model, we compute the
Bayes estimator under Stein's loss
$\hat{\Sigmabf} = E[\Sigmabf^{-1}|\ybf_1,\dots,\ybf_n]^{-1}$, where Stein's loss is
an invariant loss defined by $
L_s(\Sigmabf,\hat{\Sigmabf}) = 
tr(\Sigmabf^{-1}\hat{\Sigmabf}) - \log | \Sigmabf^{-1}\hat{\Sigmabf}| -p$.
As a reference, we also compute the sample covariance matrix (MLE).
For each scenario considered, we compute Stein's loss and summarize the 
distribution of the 25 loss values via boxplots.
All figures presented in this section 
report results on a log-scale to encourage ease of comparison across methods.

The results of the simulation study are reported in Figure \ref{sim_study_main}.
All FACE models, constructed with and without meta covariates beyond the intercept, 
more accurately estimate the true covariance than the naive sample covariance estimator in every scenario considered, by a notable margin. Additionally, in both the exchangeable and block regimes, the FACE estimator constructed with informative regime-specific meta covariates is more accurate than
the CUSP and SIS estimators.
\red{The comparison to SNE is more nuanced. In the correlated regime, SNE and FACE perform similarly. In the blocked regime, the FACE models outperform SNE for small to medium dimensions, but for $p=50$, SNE performs similarly or better than the FACE models.}
In what follows, we elaborate on the results of the simulation study for each regime.

The results of the correlated regime with population covariance $\Sigmabf_{cor}$ are plotted in Figure \ref{sim_study_main} (a).
In this regime, the intercept FACE model corresponds to an astute FACE model, since the form of the prior \red{predictive} covariance for the intercept model (Equation \ref{eqn:cmr_intercept}) corresponds to a compound symmetric covariance matrix, albeit with a greater number of unknown parameters to be estimated than is necessary for this regime. 
\red{Additionally, the prior predictive structure imposed by the SNE implementation has the correct structure of the true population covariance matrix.}
\red{Consequently,  FACE and SNE perform similarly to each other, and,}
for every dimension and sample size scenario considered, they decidedly outperform the alternative methods. 
In this non-sparse regime, the SIS and CUSP competitors seem to impose too much shrinkage towards a diagonal covariance structure.
In fact, in this regime, the CUSP estimate is outperformed even by the MLE. 
\red{In contrast, 
for a diagonal covariance regime, with zero correlation among all pairs of variables, the SIS and CUSP models that shrink the factor loading matrix towards a zero matrix outperform FACE. In this diagonal regime, all factor model priors outperform the MLE. See Appendix C for simulation results for this regime.}

The benefit of including more detailed meta covariate information is understood when analyzing the results for the block-structured covariance matrix $\Sigmabf_{block}$ in Figure \ref{sim_study_main} (b).
In this regime, results are reported for the feature aware covariance estimation model that utilizes an intercept meta covariate (FACE.I), a categorical group-identifying meta covariate (FACE.D), and a more realistic continuous meta covariate drawn from a group-specific normal distribution (FACE.C).  
In this regime, the FACE model with group-identifying meta covariates (FACE.D) outperforms \red{the MLE, SIS, and CUSP models}, often by a notable margin. 
\red{For moderately sized dimensions ($p=9,16$), FACE.D and FACE.C outperform all other methods, indicating a benefit to utilizing even imperfect outcome features within the FACE framework.
For scenarios with a large dimension ($p=50$), the SIS model has moderately larger median log loss than the FACE.D or FACE.C models, and the SNE model has moderately smaller median log loss than the FACE.D or FACE.C models.}
\red{These competitive performances may be due to the fact that there are many more open parameters in the FACE model, since, among others, the regression coefficient dimension scales with $r$.
In general, though, this highlights an overall benefit in utilizing a non-exchangeable prior in factor modeling that 
incorporates structured shrinkage.}

The continuous meta covariate used in the FACE.C model represents
imprecise group-identifying information, and, as such,
may more realistically align with meta covariates available in practice.
To this end, the benefit of incorporating auxiliary information in a modeling approach is further evidenced by the proximity of the loss for the FACE.C model to that of the FACE.D model, which utilizes group membership more directly in the meta covariate construction. 
Most notably, in the high dimension regime ($p=50$), the spreads of the losses for these two models overlap substantially in the presence of a small sample size.
Overall, the comparable performance of the FACE.C model to the FACE.D suggests there is a strong benefit to utilizing the FACE model in applications with auxiliary information on the variables being analyzed.

In total, these results suggest that feature aware meta regression is useful in improving covariance estimation, with particularly notable gains when the true covariance structure is explainable by available meta covariates. Even if no useful meta covariates are available, the naive intercept covariance regression model performs well and tends to outperform alternative popular methods, particularly in low-sample size settings.

\subsection{Matrix-variate Data Simulation Study}\label{sec:sim_matrixvariate}

In this subsection, we demonstrate the performance of the FACE model when the true covariance has a separable form.
Specifically, the population covariance matrix $\Sigmabf_{kron}$ is the Kronecker product of a $p_1$ dimensional correlation matrix with correlation 0.9 and a $p_2$ dimensional correlation matrix with correlation 0.6.
\red{As before, we examine three dimensions, $(p_1,p_2)=(2,4)$ such that $p=8$, $(p_1,p_2)=(3,5)$ such that $p=15$, and $(p_1,p_2)=(10,5)$ such that $p=50$.}
We compare the FACE model with CUSP, SIS, and the MLE under a separable normal sampling model \citep[Kron]{Dutilleul1999}.
The Kron estimate is obtained under a correctly specified parametric model and can be thought of as an oracle
procedure for this regime and, as such, is expected to outperform the other methods.

\begin{figure} 
\centering
\includegraphics[keepaspectratio,width=.8\textwidth]{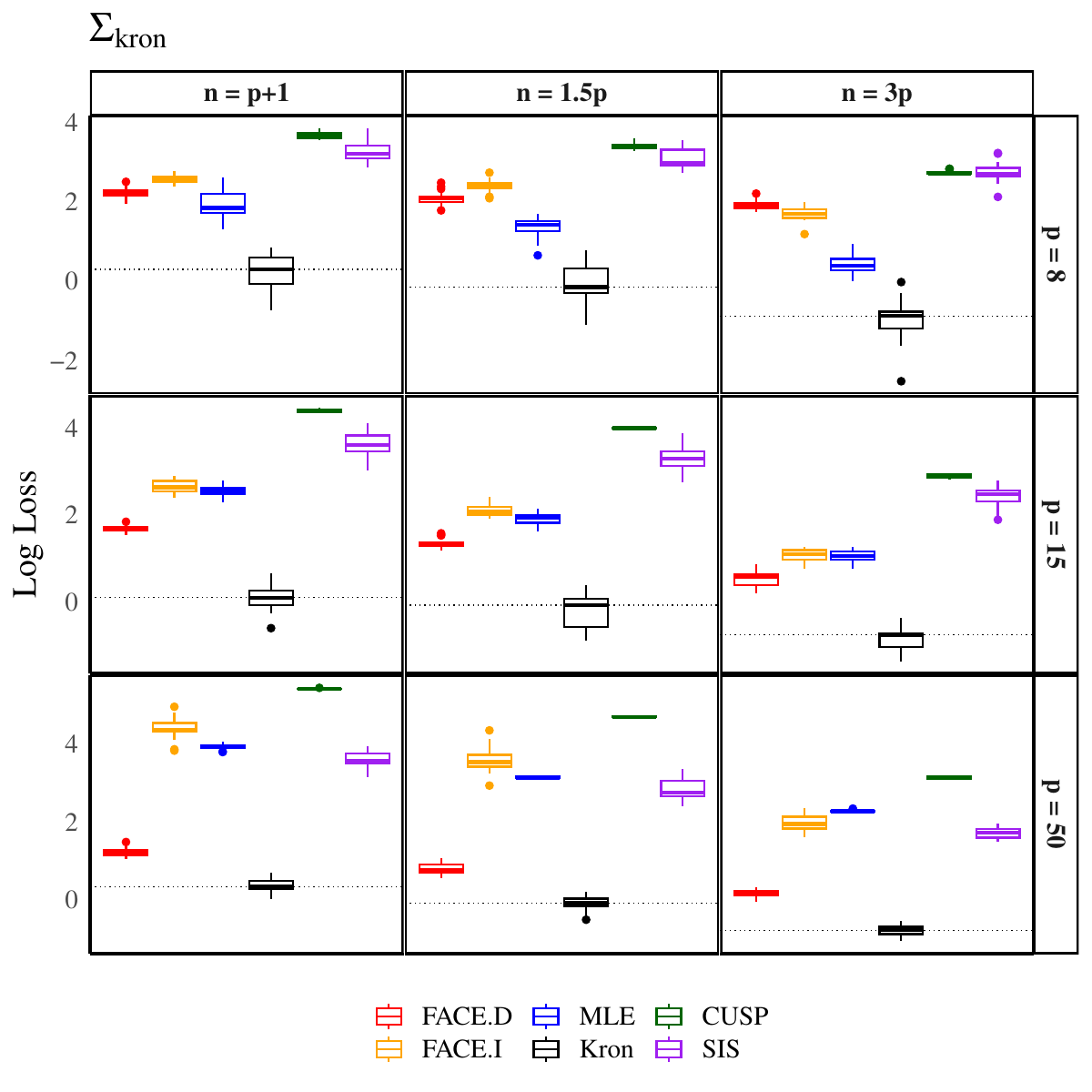}
\caption{
Stein's loss averaged over the 25 iterations for each regime, problem dimension, and sample size permutation, row-standardized by the smallest value.
\red{Results plotted for FACE with an intercept (FACE.I),
FACE with row and column membership as the outcome features (FACE.D), MLE, Kronecker MLE (KRON),
cumulative shrinkage model (CUSP), and
structured shrinkage prior with row and column membership as the outcome features (SIS).
The  dashed lines denote the minimum median loss.}
}\label{sim_output_kron}
\end{figure}

The simulation results for the regime $\Sigmabf_{kron}$ 
are reported in Figure \ref{sim_output_kron}.
As expected, the separable MLE corresponds to the smallest loss in every permutation of dimension and sample size considered. 
Overall, the FACE.D estimator performs well as the second-best option, outperforming the CUSP and SIS estimators in all the settings considered. For the small dimension setting ($p=8$), both the intercept and design matrix \red{outcome feature} FACE models outperform the competitor factor models CUSP and SIS. In larger dimensions, $p=15,50$, the FACE model using row and column identifying \red{outcome features} performs second best in every sample size setting considered, and, in the high dimension regime, there is a notably wide margin between the FACE.D loss and the next-best-performing method. In total, the results of this section indicate that the proposed FACE approach can adapt to separable structures while being flexible enough to also perform well when the separable assumption is violated.

\section{The TESIE Study: Chemical Exposure Analysis}\label{application}

Using the \red{outcome feature aware} covariance
model, we analyze the TESIE chemical exposure data. The TESIE study
collected measurements of $p = 21$ chemicals using two exposure assessment tools, household dust and silicone wristbands. Scientists are interested in identifying patterns of variation among exposures, which may be helpful in identifying vulnerable populations and
opportunities for exposure intervention. In this section, we examine the covariation of the exposures while accounting for various \red{auxiliary} properties of the measurements. Additionally, a
challenging feature of chemical exposure data is the missing values resulting from observations below a limit of detection (LOD) for chemical analyses. We show that data below the LOD can be \red{predicted} more accurately for the TESIE data by incorporating \red{outcome features}.
\red{We note that, while we focus on the task of covariance estimation by analyzing centered data,  the FACE Gibbs sampler can be trivially extended to include estimation of a mean effect. See the discussion in \S5 of \cite{Bersson2024a}, based on classical results in \cite[Theorem 3.4.4]{Mardia1979}.}

\subsection{FACE analysis: multiple \red{outcome features}}

There is a wide range of \red{outcome feature data} available for the focus chemicals of the TESIE study. Some useful information includes chemical class membership and chemical properties such as median predicted vapor pressure \citep{CompTox} and production volume \citep{EPAHPV}. As discussed earlier, chemical class information is often used in simple ways in analyses of multiple exposures, but under the assumption of conditional independence across classes. 
However, the chemical class label does not necessarily coincide with the primary use case of a chemical, so other factors should be considered when studying heterogeneity in exposure profiles across individuals. Class labels are nevertheless potentially informative about covariation in exposures, while being useful for policy and funding implications, so they should be taken into account in statistical analyses. 


To motivate the use of vapor pressure and production volume in estimating the covariance matrix among exposures, consider that
exposure levels reflect the chemical load in each household, such as, for example, how much of the chemical is applied to building materials, furnishings, consumer products, and others. Production volume may be a reasonable proxy for chemical load. In general, data on exact production volumes are limited for most chemicals, so we categorize each chemical based on its presence or absence in the High Production Volume List maintained by EPA.
Additionally, exposure levels in household dust and silicone wristbands are likely affected by how readily each chemical migrates out of products and off-gases to the indoor air or partitions to dust particles in the home. This tendency is related to the vapor pressure of a chemical. Consider, for example, two flame retardant chemicals. One can more easily off-gas from materials and be found entirely in air, and the other may never leave the product at all because of a low vapor pressure. In summary, chemicals can have drastically different behaviors, therefore differing in exposure impact, depending on these chemical properties.

Additionally, 
as discussed in Section \ref{sec:cmr_matrixvariate}, the exposures in the TESIE study have an implicit matrix structure consisting of an exposure pathway by chemical. These data consist of measurements collected from two exposure assessment tools:
dust levels, 
commonly analyzed to assess indoor environmental health exposures,
and silicone wristband levels,
a newly adopted method of assessing chemical exposures from both air and dust.
Silicone wristbands offer a promising method for passive and comprehensive data collection on chemical exposures.
In particular, 
they expand the context of the analysis as
wristband measurements reflect some behavioral attributes that dust does not, such as movement, occupational exposures, effects of clothing, time outdoors, and others.
Comparing exposure data from wristbands with other assessment tools is crucial for evaluating their effectiveness in exposure assessment. Accurately estimating the unknown covariance matrix is of particular interest in understanding, among others, how the exposures may co-vary within and across exposure sources and pathways.
For more detailed discussions of silicone wristbands as a measurement source, see \cite{Hammel2018,Wise2020}.

In summary, there are multiple possible non-overlapping categorical groupings of the exposure measurements that may be of scientific or pragmatic interest. 
In practice, often one of these groupings will be used to subset the data to be analyzed separately for each group. Typically, for example, mixtures of exposure data will be divided into subsets according to the exposure source, pathway, or chemical class, and each subset will be analyzed separately. 
This delineation is often chosen for practical reasons as the data may be thought to be too high-dimensional to analyze all exposures jointly. Moreover, it is useful for designing policy to draw conclusions using this class-information as class-specific research strategies are commonly implemented by the government \citep[e.g., ][]{EPA2021}.
Although interpretive implications are useful, 
it is not clear a priori which grouping of variables is most relevant for accurately representing the cross-exposure covariance matrix. 
These current standards for analyses may be ignoring important across-group correlations.

To resolve this ambiguity, in analyzing the TESIE data, we include all relevant \red{outcome features} using our proposed \red{feature aware} covariance meta regression framework. In particular, we model all 42 chemical exposure measurements jointly and utilize the following combination of continuous and categorical variables as \red{outcome features}: Chemical class membership, exposure assessment tool, chemical name, vapor pressure, and high production volume indicator. 
Because this analysis includes the continuous \red{outcome feature} vapor pressure,
it 
is a more flexible version 
of the FACE model discussed in Section \ref{multiple_cat_case} as it incorporates multiple categorical meta covariates, two of which represent row and column membership of what could be viewed as matrix-variate data, namely, repeated measurements of a multivariate response.

\red{Because we are incorporating a large number of meta covariates}, we proceed with estimating the 42-dimensional covariance matrix with the FACE model that incorporates a group-based \red{outcome feature regression shrinkage} mechanism.
In particular, we introduce a 
\red{group-generalized shrinkage prior 
motivated by ridge and group-lasso penalized regression}
\citep{Hoerl1970,Yuan2006}.
Specifically, for \red{outcome feature} $i\in\{1,...q\}$ belonging to category $c_i\in\{1,...,\tilde{q}\}$ for some grouping of the \red{outcome features} $\tilde{q}\leq q$, we place a \red{regression penalty} $l_{c_i}$ on variable $i$'s corresponding coefficient, to be shared with all variables belonging to group $c_i$:
\begin{align}
    \Gammabf \sim{}& N_{q\times r}(0,\Thetabf \kron \Lbf),\quad \Lbf = diag(l_{c_1},l_{c_2},\dots,l_{c_q}),  \\
    l_1,\dots,l_{\tilde{q}}\sim{}& InverseGamma (1/2,1/2). \nonumber 
\end{align}
\red{The impact of this regression shrinkage prior is considered in the out of sample prediction experiment in \S 5.2. The prediction error is notably larger for the FACE model without this penalty (red nabla in Figure \ref{TESIE_lod}).}

Details of posterior parameter estimation for the FACE procedure with this group regression \red{shrinkage} prior are contained in Algorithm 2 in Appendix D of the Supplementary Material. 
\red{In all implementations of the FACE model in the TESIE application, we set $r=p$, $(a_\theta,b_\theta) = (1/2,1/2)$, $\theta_\infty = 0.05$, and $\alpha = \lfloor p/3 \rfloor$.}
We run the proposed Gibbs sampler for 20,000 iterations, removing the first 10,000 iterations as a burn-in period and retaining every 10th iteration as a thinning mechanism.
In one implementation of the sampler using the R statistical programming
language,
20,000 iterations were completed in 19.5 minutes on a personal machine with an Apple silicone processor and 8 GB of RAM.
Convergence of the Markov chain was checked by \red{analyzing trace plots, autocorrelation, and effective sample size. Details are provided in Appendix E.}

\begin{figure} 
\centering
\includegraphics[keepaspectratio,width=.49\textwidth]{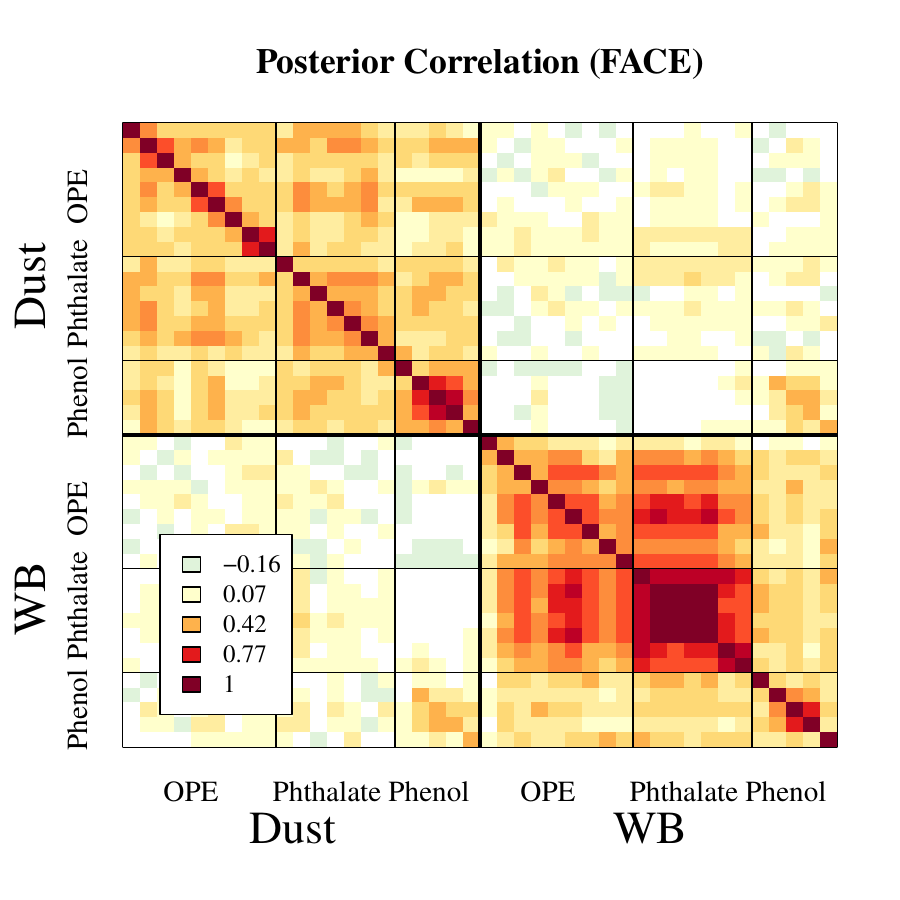}
\includegraphics[keepaspectratio,width=.49\textwidth]{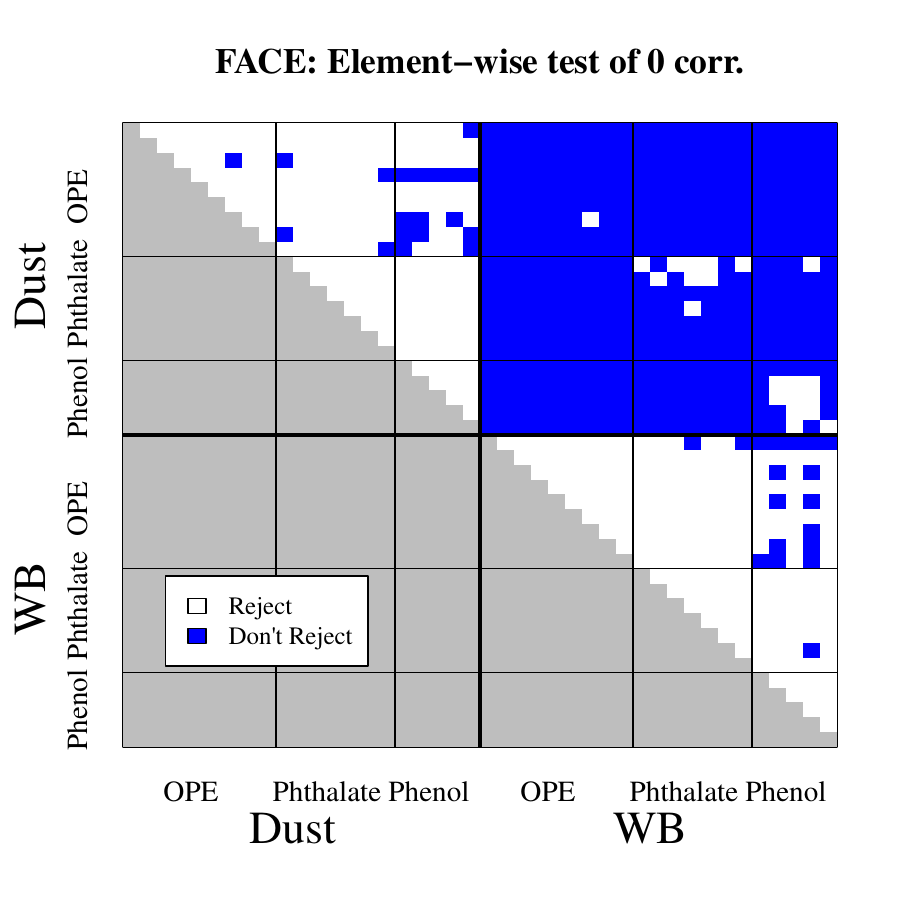} 
\caption{Analysis of the TESIE data using the FACE method. Posterior correlation (left) and inclusion of 0 in an element-wise 95\% credible interval (right). }\label{fig:CMR_TESIE}
\end{figure}

The Bayes estimate under Stein's loss of the population covariance matrix is computed from the sampled Markov chain.
For ease of visualization, the corresponding correlation matrix is plotted in the left panel of Figure \ref{fig:CMR_TESIE}.
Uncertainty quantification of the FACE covariance matrix is examined in the right panel of Figure \ref{fig:CMR_TESIE}
where 
inclusion of zero in a 95\% Bayesian credible interval based on FACE posterior analysis is represented by a blue block.
\red{Additionally, a comparison of the effective number of active factors is presented in Appendix F.}

\begin{table}[ht]
\begin{centering}
\begin{tabular}{|c||ccc|}
  \hline
 & EPB, MPB & EPB, PPB & MPB, PPB \\ 
  \hline\hline
  Within Dust & 0.71 (0.57, 0.8) & 0.66 (0.5, 0.76) & 0.83 (0.72, 0.88) \\ 
  Within WB & 0.54 (0.37, 0.69) & 0.41 (0.25, 0.59) & 0.73 (0.58, 0.81) \\ 
Between Dust/WB & 0.36 (0.14, 0.54) & 0.33 (0.12, 0.5) & 0.42 (0.22, 0.58) \\ 
  Between WB/Dust & 0.23 (0.03, 0.45) & 0.18 (-0.02, 0.4) & 0.34 (0.12, 0.51) \\ 
   \hline
\end{tabular}
\end{centering}
\caption{\red{Among phenolic parabans, posterior estimates of correlation with corresponding 95\% credible interval bounds in parenthesis. Elements in the row labeled `Within Dust' correspond to correlation between the chemicals listed in the first rows among exposures from dust samples (upper left quadrant of the correlation matrix in Figure \ref{fig:CMR_TESIE}). Likewise for the `Within WB' row. Elements in the row labeled `Between Dust/WB'  correspond to correlations between the first labeled chemical measured by dust and the second labeled chemical measured by wristband. For example, the estimated posterior correlation between EPB exposure measured by dust and MPB exposure measured by WB is 0.36. Likewise for the `Between WB/Dust' row.}}\label{table:paraban_highcor}
\end{table}

The exposure correlation matrix exhibits many \red{credibly non-zero} and informative dependencies, \red{that is, correlations with credible intervals that do not contain zero}, between chemicals and exposure assessment tools. In general, many of the chemicals analyzed are widely prevalent, so
positive correlations are expected within the assessment tool. For one, strong positive correlations are exhibited among exposures to chemicals that are often used in products in combination. For example, the three parabens EPB, MPB, and PPB are often used in combination in personal care products, so we expect positive correlations among exposures to these chemicals. For these three chemicals, strong positive correlation is exhibited in both within-exposure matrix correlations and between-exposure matrix correlations, \red{see Table \ref{table:paraban_highcor}}.
Similarly, 2IPPDPP, 24DIPPDPP, and B4TBPP, 
OPEs commonly used in the Firemaster 550 and Firemaster 600 flame retardant mixtures \citep{Phillips2017}, are correlated, particularly in silicone wristbands \red{(Table \ref{table:ope_highcor})}.

\begin{table}[ht]
\begin{centering}
\begin{tabular}{|c||ccc|}
  \hline
 & 24DIPPDPP, 2IPPDPP & 24DIPPDPP, B4TBPP & 2IPPDPP, B4TBPP \\ 
  \hline\hline
Within Dust & 0.5 (0.33, 0.66) & 0.35 (0.19, 0.54) & 0.34 (0.19, 0.56) \\ 
  Within WB & 0.65 (0.5, 0.74) & 0.49 (0.34, 0.63) & 0.57 (0.4, 0.69) \\ 
   \hline
\end{tabular}
\end{centering}
\caption{\red{Among flame retardant OPEs, posterior estimates of correlation with corresponding 95\% credible interval bounds in parenthesis. Elements in the row labeled `Within Dust' correspond to correlation between the chemicals listed in the first rows among exposures from dust samples (upper left quadrant of the correlation matrix in Figure \ref{fig:CMR_TESIE}). Likewise for the `Within WB' row.}}\label{table:ope_highcor}
\end{table}

In addition to identifying some expected strong positive correlations among exposures, our analysis identifies less-anticipated correlations. One such example is that phthalates DEP, DiBP, BBP, DBP, and DEHP exhibit 
\red{credibly}
positive correlations between the exposure assessment tools \red{(Table \ref{table:phthalate_highcor})}. Although these phthalates share some overlapping uses, they are also found in distinct products, which may indicate that certain groups of people experience higher exposure levels. This has important implications for studies
examining the health impacts of phthalate exposure, as it underscores the need to consider exposure patterns in multiple compounds. Furthermore, TCS, an antimicrobial chemical, showed a
strong correlation between exposure assessment tools, suggesting that silicone wristbands effectively capture TCS exposure from household dust. This has important implications for exposure assessment, as wristbands provide a less invasive and more resource efficient
alternative to collecting household dust samples.

\begin{table}[ht]
\begin{centering}
\begin{tabular}{|c||ccccc|}
  \hline
 & DEP, BBP & DEP, DBP & DEP, DEHP & DiBP, DBP & DiBP, DEHP \\ 
  \hline\hline
Between Dust/WB & 0.21 (0.03, 0.43) & 0.2 (0.02, 0.43) & 0.19 (0, 0.43) & 0.3 (0.1, 0.51) & 0.24 (0.04, 0.47) \\ 
   \hline
\end{tabular}
\end{centering}
\caption{\red{Among phthalates, posterior estimates of correlation with corresponding 95\% credible interval bounds in parenthesis. Elements in the row labeled `Between Dust/WB'  correspond to correlations between the first labeled chemical measured by dust and the second labeled chemical measured by wristband. For example, the estimated posterior correlation between DEP exposure measured by dust and BBP exposure measured by WB is 0.21. }}\label{table:phthalate_highcor}
\end{table}

The FACE covariance estimate is much less shrunk toward a diagonal matrix than the estimate obtained from the SIS approach, which also uses \red{outcome features} \red{(Figure \ref{fig:SIS_TESIE})}. Moreover, with the FACE approach, there are many more 
non-zero correlations that are deemed \red{credible}
than with the naive or SIS analyses. 
In addition, complex dependencies among exposures are evident in the FACE posterior analysis. For one, there is a nearly separable across-pathway effect of phenolic/paraben compounds \red{with a credible inteval that does not contain zero}. This is evidenced by the shared pattern in the phenol by phenol covariances both within
and across exposure assessment tools.
This suggests an informative within-phenol exposure pattern that is shared across exposure pathways.

There are some similarities in conclusions from the FACE and SIS analyses. The FACE and SIS estimates exhibit some similar patterns of strong positive correlation, particularly among phthalate exposures measured from the same exposure pathway. Furthermore, all three estimates exhibit strong positive correlations among three phenols EPB, MPB, and PPB. In fact, these chemicals are parabans that tend to be used in combination in personal care products, so this strong positive correlation is expected. All \red{credible} discoveries (white blocks) found using the classical method are also found using the FACE approach. A similar conclusion holds for discoveries found via the SIS approach. However, the FACE approach produces significantly more discoveries of non-zero correlations. This is particularly notable since
the FACE method is based on a hierarchical model, and, as such, incorporates an automatic Bayesian multiplicity adjustment \cite[p. ~96]{Gelman2014}.

\subsection{\red{Predicting} values below LOD}

\begin{figure}
    \centering \includegraphics[width=0.45\linewidth]{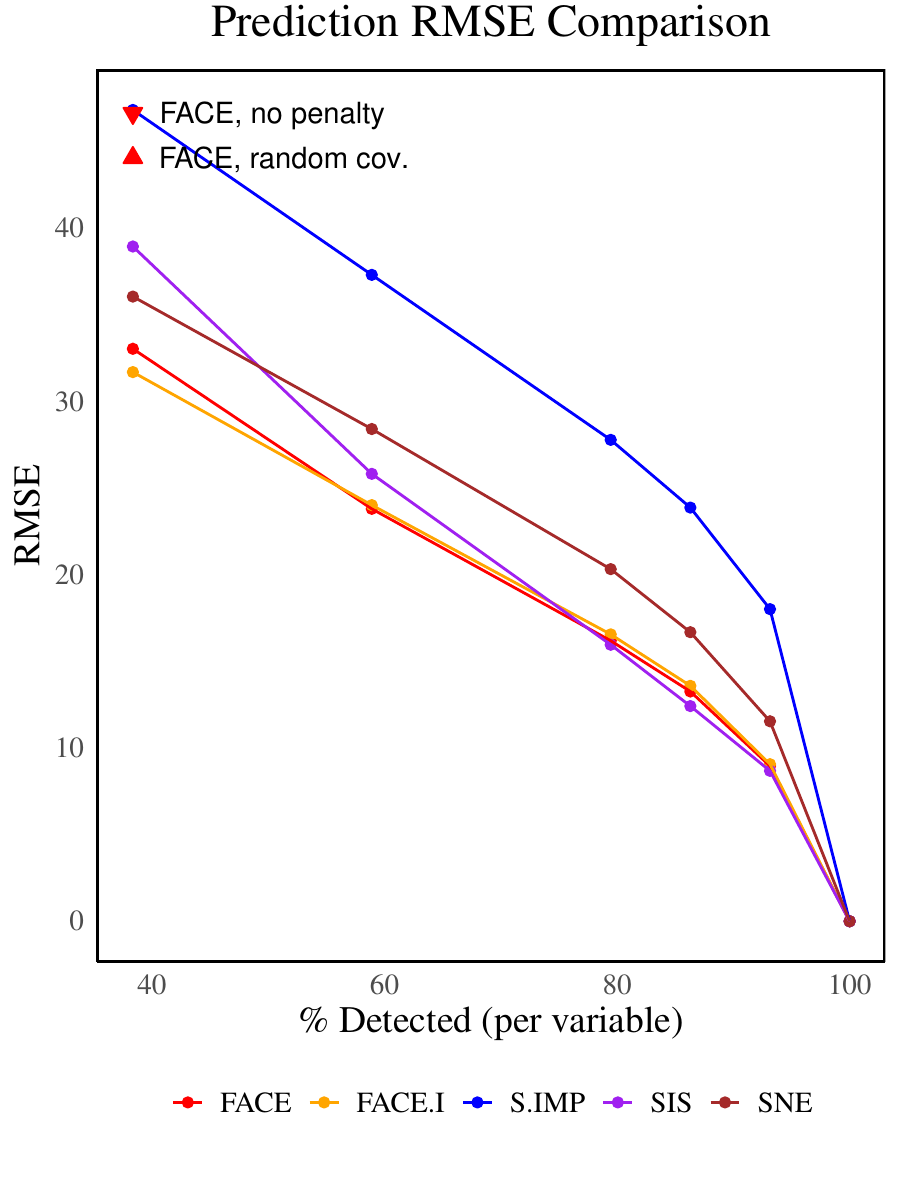}
    \caption{Square root of the mean squared error (RMSE) of \red{predicted} values.
    \red{Results plotted for FACE with an intercept (FACE.I),
    FACE with application-specific outcome features (FACE),
    single imputation (S.IMP),
    structured shrinkage prior with application-specific outcome features (SIS), and the structured matrix normal prior for an exchangeable matrix (SNE). 
    Results plotted for FACE with no regression variable shrinkage (red nabla, "FACE, no penalty") and FACE with randomly generated outcome features (red triangle, "FACE, random cov.").}
    }
    \label{TESIE_lod}
\end{figure}

To understand the impact of various analysis procedures on the accuracy of \red{prediction} of data below a limit of detection (LOD), we perform an experiment on TESIE data. We hold out the smallest $n_{test}$ values from each column as test data points and \red{predict} them based on an experimental limit of detection that is the average of the largest value in the test set and the smallest value in the training set. 
We \red{predict} these missing values within the Gibbs samplers for the FACE model with just an intercept \red{outcome feature} (FACE.I), the FACE model with the combination of categorical and continuous \red{outcome features} (FACE), the SIS model with the same \red{outcome features} used in the FACE model, and 
\red{the structured matrix normal prior for an exchangeable matrix (SNE)}. 
We estimate each value with the posterior mean from the resulting MCMC samples, and compute the root mean squared error (RMSE) of the testing data samples. 
Additionally, we compare the results to the RMSE obtained from a single imputation that sets each missing value to the LOD/$\sqrt{2}$. 
\red{Furthermore, to evaluate robustness of the FACE model, 
for the lowest percent detected considered, we compute RMSE for FACE with no regression variable shrinkage and 
FACE with randomly generated outcome features.
}

The single imputation is overwhelmingly the most common approach for imputing values below a limit of detection in practice \citep{Helsel2006,Venier2016}.
As such, it is of primary interest to compare the imputation accuracy of the state-of-the-art approaches to this baseline naive approach. Moreover, inclusion of variables in an analysis is commonly determined based on percent detected, such that variables with low levels of detection are excluded \citep{Hites2019}. To this end,
we are particularly interested in how the
accuracy of the imputation varies as a function of the percent of measurements detected above the LOD, $100(n-n_{test})/n\%$ ("\% Detected").

The results for varying numbers of samples below the limit of detection within each column are plotted in Figure \ref{TESIE_lod}. For all values of $n_{test}$ considered, the \red{FACE and FACE.I} models result in more accurate \red{prediction} than the single imputation approach. In particular, even for high levels of detection, \red{prediction} error is notably improved over the naive method. The \red{prediction} based on the SIS model varies in performance depending on the percent detected. When a small percentage of the data is below the detection limit, the \red{prediction} from the SIS model performs similarly to or slightly better than the FACE models. However, as the percent detected decreases, the \red{prediction} based on the SIS model is notably less accurate than that from FACE models and can even be worse than that from the naive single imputation approach. 
\red{The SNE model returns worse predictions than either the FACE or FACE.I model for each detection level we considered, and it is worse than the SIS model for all but the most sparse percent detected considered.}
The improved \red{prediction} accuracy of the FACE approach can in turn improve analyses of exposure profiles.

\red{To
explicate the sensitivity to the
informativeness of the outcome features used in the FACE model,
we run the prediction experiment using randomly generated outcome features
for the lowest percent of data detected. 
Specifically,
to mimic the real outcome features used as covariates,
we construct 4 random categorical features, one binary feature generated from a Bernoulli distribution with success probability of 0.6, and one continuous feature generated from a standard normal distribution.
The RMSE for this experiment is marked by a red triangle in Figure \ref{TESIE_lod}.
Using these independent and therefore noninformative feature covariates results in prediction accuracy similar to a naive single imputation. 
As this is a real data experiment, the real outcome features used in the FACE model may be considered imperfect. 
Therefore,
FACE with real outcome features outperforming FACE with faux outcome features supports the conclusion that the model can leverage 
dependencies from auxiliary information.  
}

\section{Discussion}\label{discussion}

In this work, we propose a feature aware covariance estimation prior that allows for improved covariance estimation for high dimensional regimes by integrating auxiliary information on the variables into a model-based approach. Our method utilizes a latent factor model framework that allows flexibility in the rank of the squared factor loading matrix. Furthermore, the FACE method enables structured shrinkage of a covariance matrix towards a non-diagonal form. We show how various \red{outcome feature} data types can allow for shrinkage towards a wide array of commonly utilized structures such as compound symmetric and block symmetric covariance matrices. Moreover, parameter estimation, inferences, and \red{prediction} of missing data, including under a limit of detection, are straightforward for our feature aware covariance model using a Gibbs sampler. 

Using \red{feature aware covariance estimation} to analyze data from the TESIE study yields an informative estimate of the correlation pattern among a wide array of exposures measured from two different exposure assessment tools. In particular, there are strong positive correlations between chemicals from different chemical classes and assessment tools, leading to new insights into exposure patterns. Moreover, 
we show that the proposed FACE model can reduce the error in imputing exposures under the limit of detection, based on an experiment using TESIE data. This promising result highlights the need for and benefit of using realistic covariance structure models for imputation instead of naive methods.

\red{
Inference that incorporates outcome features within the FACE framework is shown to be sufficiently robust and results in more accurate estimation in simulation and more accurate out of sample prediction in the data application.
In the outcome feature regression on each factor loading vector, we utilize a linear regression equation, however,  non-linear functional forms can be captured with, for example,
basis expansions of covariates, among other standard extensions of ordinary linear regression. 
Future work exploring the benefit of incorporating additional flexibility 
in this way, or integrating yet more flexibility with, for instance, a Gaussian process, would be beneficial.
}

There are a few possible extensions of the FACE framework that may prove useful, particularly for environmental health applications. For one, understanding how patterns of exposure covariation differ for different at-risk groups in the population is of particular interest. As the FACE approach allows scientists to better assess true correlations and associations with small sample sizes, extending the framework to account for multiple sub-populations is of immediate interest. Separately, inclusion of a more interpretable variable selection mechanism on the \red{outcome features} could allow for useful insight into the importance of varying available auxiliary data. 
\red{In the TESIE application, 
there are many possible outcome features that 
could be considered.
To demonstrate possible curation of outcome features for an analysis, we incorporate a large number of outcome features obtained from various sources. 
Some of these features, like chemical class, were obtained from the original data set, and others, like vapor pressure, were mined from reputable online data sources.
}
Understanding which of these chemical properties are most impactful in estimating the covariance matrix could lead to a better understanding of exposures.

More generally, in this work, we elaborate on the perspective that additional information on the variables being analyzed may be encoded as \red{outcome features}, and formally including this information in a modeling framework can yield improved accuracy of model parameters over standard approaches.
To this end, in Section \ref{sec:cmr_matrixvariate} we highlight a potential new framework for covariance estimation for matrix-variate data that encodes the membership of the rows and columns as \red{outcome features}, and the benefit of this approach in improving the accuracy of covariance estimation is conveyed in the simulation (Section \ref{sec:sim_matrixvariate}).
With this motivation, an interesting avenue for future work may explore how the FACE framework may be used to formulate alternative modeling approaches for structured data beyond matrix-variate.

\begin{funding}
This work was supported by the NIH National Institute of
Environmental Health Sciences grant
P42-ES010356.
David Dunson was supported by funding from the United States National Institutes of Health R01-ES035625.
\end{funding}

\begin{supplement}
\stitle{Appendix A: Chemical Details}
\sdescription{A table containing details of the chemicals measured in the TESIE study including chemical acronym, full chemical name, chemical class, and primary use of the chemical.}
\end{supplement}
\begin{supplement}
\stitle{Appendix B: Derivation of the Prior \red{Predictive} Covariance}
\sdescription{Derivation of the form of the prior \red{predictive} covariance matrix for the feature aware covariance model.}
\end{supplement}
\begin{supplement}
\stitle{\red{Appendix C: Additional Simulation Analyses}}
\sdescription{\red{Simulation study results for a regime with an identity true population covariance, $\Sigmabf_I = \eye$.}}
\end{supplement}
\begin{supplement}
\stitle{Appendix D: Group Generalized Variable \red{Shrinkage}}
\sdescription{A Gibbs sampler for the feature aware covariance model with a variable selection prior on the \red{outcome feature} regression.}
\end{supplement}
\begin{supplement}
\stitle{\red{Appendix E: TESIE Application MCMC Convergence Analysis}}
\sdescription{\red{Analysis of convergence of the Monte Carlo Markov chain for the TESIE data application.}}
\end{supplement}
\begin{supplement}
\stitle{\red{Appendix F: TESIE Application Comparison of Effective Latent Dimension}}
\sdescription{\red{
Posterior estimates of the effective latent dimension for the FACE.I, FACE, SIS, and SNE methods.}}
\end{supplement}

\bibliographystyle{imsart-nameyear} 
\bibliography{misc/library}       

\end{document}